\documentclass[useAMS,usegraphicx,usenatbib]{mn2e}
\usepackage{color}
\usepackage{wasysym}
\usepackage{url}
\usepackage{multirow}

\def\kms{km\,s$^{-1}$}
\def\Ha{H$\alpha$}

\def\CII{C\,{\sc ii}}

\def\SiI{Si\,{\sc i}}
\def\SiII{Si\,{\sc ii}}

\def\CaII{Ca\,{\sc ii}}

\def\FeII{Fe\,{\sc ii}}
\def\FeIII{Fe\,{\sc iii}}
\def\CoII{Co\,{\sc ii}}
\def\CoIII{Co\,{\sc iii}}
\def\NiII{Ni\,{\sc ii}}

\def\CaII{Ca\,{\sc ii}}

\def\Nifs{$^{56}$Ni}
\def\Cofs{$^{56}$Co}
\def\Fefs{$^{56}$Fe}
\def\Msun{M$_{\odot}$}

\def\ebv{$E(B-V)$}
\def\dm15{$\Delta m_{15}(B)$}
\def\MCh{M$_\mathrm{Ch}$}
\def\lesssim{\mathrel{\hbox{\rlap{\hbox{\lower4pt\hbox{$\sim$}}}\hbox{$<$}}}}
\let\la=\lesssim
\def\gtrsim{\mathrel{\hbox{\rlap{\hbox{\lower4pt\hbox{$\sim$}}}\hbox{$>$}}}}
\let\ga=\gtrsim

\def\aj{AJ}%
\def\apj{ApJ}%
\def\apjl{ApJ}%
\def\aap{A\&A}%
\def\mnras{MNRAS}%
\def\pasp{PASP}%
\def\nat{Nature}%
\title[`Super-Chandrasekhar' SNe Ia at nebular epochs]{`Super-Chandrasekhar' 
Type Ia Supernovae at nebular epochs\thanks{Based on observations at ESO Paranal, 
Prog. 281.D-5043, 083.D-0728 and 085.D-0701.}
}
\author[Taubenberger et al.]{S. Taubenberger$^{1}$
\thanks{E-mail: tauben@mpa-garching.mpg.de},
M.~Kromer$^{1}$, S.~Hachinger$^{2,3}$, P.~A.~Mazzali$^{1,3}$,  S.~Benetti$^{3}$,
\newauthor P.~E.~Nugent$^{4,5}$, R.~A.~Scalzo$^{6}$, R.~Pakmor$^{7}$, V.~Stanishev$^{8}$, J.~Spyromilio$^{9}$, 
\newauthor F.~Bufano$^{10}$, S.~A.~Sim$^{6}$, B.~Leibundgut$^{9}$ \& W.~Hillebrandt$^{1}$\\
$^{1}$Max-Planck-Institut f\"{u}r Astrophysik, Karl-Schwarzschild-Str. 1, 85741 Garching bei M\"{u}nchen, Germany\\
$^{2}$Julius-Maximilians-Universit\"{a}t W\"{u}rzburg, Emil-Fischer-Str.31, 97074 W\"{u}rzburg, Germany\\
$^{3}$INAF Osservatorio Astronomico di Padova, Vicolo dell'Osservatorio 5, 35122 Padova, Italy\\
$^{4}$Lawrence Berkeley National Laboratory, Berkeley, CA, 94720, USA\\
$^{5}$Department of Astronomy, University of California, Berkeley, CA, 94720-3411, USA\\
$^{6}$Research School of Astronomy \& Astrophysics, Mount Stromlo Observatory, Cotter Road, Weston ACT 2611, Australia\\
$^{7}$Heidelberger Institut f\"{u}r Theoretische Studien, Schloss-Wolfsbrunnenweg 35, 69118 Heidelberg, Germany\\
$^{8}$CENTRA -- Centro Multidisciplinar de Astrof\'{i}sica, Instituto Superior T\'{e}cnico, Av. Rovisco Pais 1, 1049-001 Lisbon, Portugal\\
$^{9}$European Southern Observatory (ESO), Karl-Schwarzschild-Str. 2, 85748 Garching bei M\"{u}nchen, Germany\\
$^{10}$Departamento de Ciencias Fisicas, Universidad Andres Bello, Avda. Republica 252, Santiago, Chile}

\begin{document}

\date{Accepted 2013 April 16.  Received 2013 April 16; in original form 2012 December 18.}

\pagerange{\pageref{firstpage}--\pageref{lastpage}} \pubyear{2013}

\maketitle

\label{firstpage}

\begin{abstract}
We present a first systematic comparison of superluminous Type Ia 
supernovae (SNe~Ia) at late epochs, including previously unpublished 
photometric and spectroscopic observations of SN~2007if, SN~2009dc 
and SNF20080723-012. 
Photometrically, the objects of our sample show a diverse late-time 
behaviour, some of them fading quite rapidly after a light-curve 
break at $\sim$\,150--200\,d. The latter is likely the result of flux 
redistribution into the infrared, possibly caused by dust formation, 
rather than a true bolometric effect. Nebular spectra of superluminous 
SNe~Ia are characterised by weak or absent [\FeIII] emission, pointing 
at a low ejecta ionisation state as a result of high densities. 
To constrain the ejecta and \Nifs\ masses of 
superluminous SNe~Ia, we compare the observed bolometric light curve 
of SN~2009dc with synthetic model light curves, focusing on the 
radioactive tail after $\sim$\,60\,d. Models with enough \Nifs\ to 
explain the light-curve peak by radioactive decay, and at the same time 
sufficient mass to keep the ejecta velocities low, fail to reproduce 
the observed light-curve tail of SN~2009dc because of too much 
$\gamma$-ray trapping. We instead propose a model with $\sim$\,1 
\Msun\ of \Nifs\ and $\sim$\,2 \Msun\ of ejecta, which may be 
interpreted as the explosion of a Chandrasekhar-mass white dwarf (WD)
enshrouded by 0.6--0.7 \Msun\ of C/O-rich material, as it could 
result from a merger of two massive C/O WDs. 
This model reproduces the late light curve of SN~2009dc well. 
A flux deficit at peak may be compensated by light 
from the interaction of the ejecta with the surrounding material.
\end{abstract}


\section{Introduction}
\label{Introduction}

Type Ia supernovae (SNe~Ia) are powerful distance indicators and have been used 
to infer the accelerating expansion of the Universe \citep{riess1998a,schmidt1998a,
perlmutter1999a}. They are considered amongst the most promising tools 
to distinguish between a Cosmological Constant and other forms of Dark Energy 
\citep{goobar2011a}. Normal SNe~Ia \citep{branch1993a} are excellently suited 
for this purpose owing to their remarkable homogeneity in peak luminosity and 
light-curve shape. However, not all SNe~Ia are so `well-behaved'. Several 
subclasses \citep[e.g.][]{filippenko1992b,leibundgut1993a,li2003a,howell2006a,
foley2010a,ganeshalingam2012a} are known to defy normalisation through the 
usual relations between light-curve width and peak luminosity \citep{phillips1993a,
phillips1999a}, one of them being the group of superluminous Type Ia SNe.

Compared to normal SNe~Ia, superluminous SNe~Ia are characterised by a bright 
light-curve peak, a slow light-curve evolution during the photospheric phase 
and moderately low ejecta velocities \citep{howell2006a,branch2006b,hicken2007a,
yamanaka2009a,scalzo2010a,yuan2010a,silverman2011a,taubenberger2011a}.  
Modelling suggests ejecta masses far in excess of the Chandrasekhar-mass 
(M$_\mathrm{Ch}$) limit of non-rotating white dwarfs (WDs) and the production 
of about 1.5 M$_\odot$ of $^{56}$Ni, precluding the interpretation of these 
events as thermonuclear explosions of Chandrasekhar-mass (\MCh) WDs. For this 
reason they are commonly referred to as `super-Chandrasekhar' SNe~Ia in the 
literature \citep{howell2006a,branch2006a}. 

Models of thermonuclear explosions are severely 
challenged by superluminous SNe~Ia. Proposed explanations range from the 
explosions of differentially rotating massive WDs \citep[e.g.,][]{howell2006a} 
to WD mergers \citep{hicken2007a} with possible interaction between the 
actual SN ejecta and surrounding circumstellar medium (CSM) left by the merger 
\citep{scalzo2010a,scalzo2012a,hachinger2012a}, and thermonuclear explosions in 
the degenerate cores of AGB stars \citep{taubenberger2011a}. However, none of 
these models is without problems. In particular, there is no evidence of 
ejecta--CSM interaction in the form of narrow emission lines in the photospheric 
spectra of any of these SNe. The problems to explain superluminous SNe~Ia 
with thermonuclear explosions even led to the speculation that these objects 
might instead have a core-collapse origin, with the early light curve 
possibly powered by magnetar heating \citep{taubenberger2011a}.

Great power to discriminate different explosion scenarios has traditionally 
been ascribed to nebular spectra. Sampling also the inner ejecta, differences 
between a thermonuclear explosion and a core-collapse event should become 
evident in nebular spectra, given the entirely different nucleosynthetic 
footprint. Moreover, from late-time light curves a refined estimate of 
the \Nifs\ mass can be made, complementing the estimates based on the 
light-curve peak.

So far, only few late-time observations of superluminous SNe~Ia have been 
published. For SNe~2007if \citep{scalzo2010a,yuan2010a} and 2009dc 
\citep{yamanaka2009a,tanaka2010a,silverman2011a,taubenberger2011a} nebular 
spectra were shown by \citet{yuan2010a,silverman2011a} and 
\citet{taubenberger2011a}. These spectra were dominated by [\FeII] emission 
lines, [\FeIII] features were quite weak. For SN~2006gz \citep{hicken2007a} 
\citet{maeda2009a} presented a nebular spectrum that seemed to lack 
completely the prominent [\FeII] and [\FeIII] emission lines shortward of 
5500\,\AA, normally the hallmark features of SNe~Ia at late epochs. At the 
same time, SN~2006gz was unexpectedly dim one year after maximum light, 
indicating an enhanced fading after the photospheric phase. A similar 
trend, though less extreme, was reported for SN~2009dc after $\sim$\,200\,d 
\citep{silverman2011a,taubenberger2011a}. 

In this work additional late-time data of the superluminous SNe~Ia SN~2007if, 
SN~2009dc and SNF20080723-012 \citep{scalzo2012a} are presented and 
analysed. Complemented by the published data of SN~2006gz 
\citep{maeda2009a}, a first systematic comparison of these objects 
during the nebular phase is performed, revealing significant diversity 
in the spectral appearance and the luminosity and decline rate of the 
radioactive tail of the light curve. The paper is organised as follows. 
In Section~\ref{Observations and data reduction} the methods used to 
reduce and calibrate the new data are described. Section~\ref{Luminosity 
evolution} presents the late-time luminosity evolution, 
Section~\ref{Spectroscopic comparison} the properties and peculiarities 
of the nebular spectra. In Section~\ref{Discussion} the findings are 
discussed, synthetic bolometric light curves for a set of models are 
compared to the observations, and an attempt is made to propose a uniform 
model for at least some superluminous SNe~Ia. A summary and conclusions 
are given in Section~\ref{Conclusions}.

\section{Observations and data reduction}
\label{Observations and data reduction}

\begin{table*}
\caption{Spectra of SNe~2007if, SNF20080723-012 and SN~2009dc.}
\label{spectra} 
\begin{footnotesize}
\begin{center}
\begin{tabular}{@{}llcrccccll@{}}
\hline
UT date & MJD & SN & Epoch$^a$ & Telescope/Instrument & Setup & Exposure time \\
\hline
2008 Sep 24 & 54\,733.2 & SN~2007if & 358.2 & VLT\,+\,FORS2    & 300V        & 2850\,s $\times$ 2 \\
2008 Oct 22 & 54\,761.2 & SN~2007if & 384.3 & VLT\,+\,FORS2    & 300V        & 2850\,s $\times$ 3 \\
2009 Apr 26 & 54\,947.3 & SNF       & 247.9 & VLT\,+\,FORS2    & 300V        & 2880\,s $\times$ 4 \\
2009 May 22 & 54\,973.9 & SNF       & 272.7 & VLT\,+\,FORS2    & 300V        & 2850\,s $\times$ 3 \\
2009 Jun 19 & 55\,001.2 & SNF       & 298.1 & VLT\,+\,FORS2    & 300V        & 2850\,s $\times$ 3 \\
2010 May 10 & 55\,326.2 & SN~2009dc & 371.6 & VLT\,+\,XShooter & UVB,VIS,NIR & 1350\,s $\times$ 4 \\
\hline
\end{tabular}
\\[1.5ex]
\flushleft
$^a$~Phase in rest-frame days with respect to $B$-band maximum [MJD $= 54\,348.4$ 
for SN~2007if \citep{scalzo2010a}, MJD $= 54\,680.9$ for SNF20080723-012 \citep{scalzo2012a}, 
MJD = $54\,946.6$ for SN~2009dc \citep{taubenberger2011a}].\\
\end{center}
\end{footnotesize}
\end{table*}

The late-time data presented here were obtained with the FORS2 instrument 
mounted at the 8.2\,m Very Large Telescope (VLT) UT-1, with XShooter at 
the 8.2\,m VLT UT-2, and with CAFOS at the 2.2\,m Telescope of the Calar 
Alto Observatory. The CAFOS and FORS2 data were reduced following 
standard procedures within IRAF, including bias subtraction and flatfielding. 
An optimal, variance-weighted extraction \citep*{horne1986a} of the spectra 
(Table~\ref{spectra}) was performed, and arc-lamp exposures were used to 
determine the dispersion solution. The XShooter observations 
(Table~\ref{spectra}) were 
pipeline-processed\footnote{http:/$\!$/www.eso.org/observing/dfo/quality/XShooter/\\ 
\hspace*{0.28cm}pipeline/pipe\_gen.html} to create linearised, sky-subtracted, 
wavelength-calibrated 2D spectra (for each of the UVB, VIS and NIR channels) 
out of the curved Echelle orders of XShooter. The 2D spectra were then 
optimally extracted. Telluric-feature removal and a rough flux calibration were 
accomplished using observations of spectrophotometric (UVB, VIS) or telluric 
(NIR) standard stars. The flux calibration of all optical spectra was checked 
against the photometry and adjusted when necessary. Lacking contemporaneous 
photometry, this was not possible for the XShooter NIR spectrum, which was 
instead scaled to match the VIS spectrum in the region of overlap. Since this 
region is small, noisy and affected by the steep transmission edge of the 
dichroic, the obtained calibration is estimated to be no more precise than 
to a factor of $\sim$\,2.

\subsection{Host-galaxy subtraction}
\label{Host-galaxy subtraction}

\subsubsection{SNF20080723-012}
\label{SNF20080723-012}

\begin{figure}
   \centering
   \includegraphics[width=8.4cm]{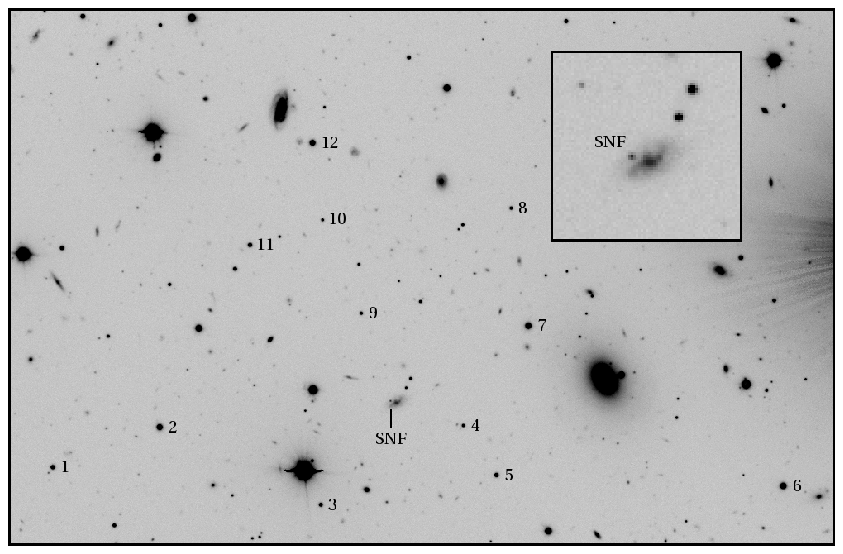}\\
   \vspace*{0.3cm}
   \includegraphics[width=8.4cm]{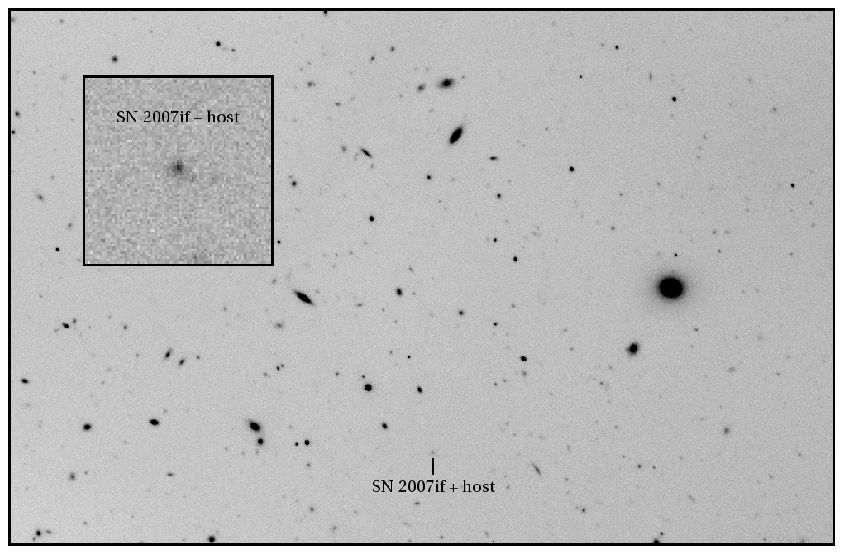}
   \caption{$V$-band image of the SNF20080723-012 field (top panel) and $I$-band 
   image of the SN~2007if field (bottom panel), taken with the VLT UT-1 + FORS2 
   on UT 2009 May 22 and 2008 September 24, respectively. 
   The fields of view are $4.2 \times 2.7$ arcmin$^2$, and the regions around the 
   SNe are triply enlarged in the insets. North is up and east to the left. The 
   local sequence stars around SNF20080723-012 (Table~\ref{std}) are indicated.}
   \label{fig:chart}
\end{figure}

SNF20080723-012 was located within a spiral host galaxy (Fig.~\ref{fig:chart}, 
top). Luckily, it was sufficiently off-set from the bright spiral arms that a 
direct point-spread-function (PSF) fitting measurement was feasible under very 
good seeing conditions. Such conditions ($\sim$\,0.5 arcsec FWHM) were met on 
2009 May 22 with the VLT + FORS2. In the images of that night a direct 
PSF-fitting measurement of the SN magnitudes was performed. At the same time, 
the PSF-subtracted images were used as pseudo-templates to perform a host-galaxy 
subtraction at earlier epochs when the seeing conditions were poorer. The 
possible error introduced by residual SN flux in the pseudo-templates is 
estimated to be small ($\la 0.1$ mag) for the VLT photometry of 2009 April 26 
and negligible for the earlier Calar Alto photometry.

\subsubsection{SN 2007if}
\label{SN 2007if}

Background contamination is a more severe problem in the late-time 
observations of SN~2007if. Its dwarf host galaxy at redshift $z = 0.074$ 
has an almost stellar PSF, and the SN has no discernible offset 
(Fig.~\ref{fig:chart}, bottom; see also \citealt{childress2011a}). By the 
time of our observations, the SN had faded below the brightness of the host. 
Accordingly, PSF-fitting magnitudes were dominated by host-galaxy light, and 
the extracted spectra showed clear evidence of an underlying continuum. 

The continuum in the spectra was removed by rescaling and subtracting a 
spectral template of the host galaxy as presented by \citet{childress2011a}. 
In doing so, we assumed near zero flux in those spectral regions where 
emission lines from the SN are not expected.

Since no sufficiently deep imaging templates were available in our photometric 
bands to perform a conventional template subtraction, we applied what may be 
called a numerical host subtraction. To this aim, we derived synthetic host 
magnitudes in the $BV\!RI$ bands from the host-galaxy spectral template of 
\citet{childress2011a}, scaled to match the $g$-band magnitude of the 07if host 
as reported by the same authors and \citet{scalzo2010a}. We then measured 
the `SN + host' magnitudes in our VLT images using aperture photometry with 
a 2.5 arcsec radius, large enough to include virtually all the light from the 
host (and the high-$z$ background galaxy reported by \citealt{childress2011a}). 
Subtraction of the synthetic host fluxes from the `SN + host' fluxes finally 
yielded SN magnitudes in each band.\footnote{Note that \citet{yuan2010a} obtained
a somewhat fainter host $g$-band magnitude than \citet{scalzo2010a} and 
\citet{childress2011a}, which would result in brighter SN magnitudes. The 
difference between the two numbers has been taken into account in the error 
assigned to the host magnitude, and propagated to the error of the SN 
magnitudes.}

\begin{table*}
\caption{$S$- and $K$-corrected photometry of SN~2007if and SNF20080723-012.}
\label{mags} 
\begin{footnotesize}
\begin{center}
\begin{tabular}{@{}llcrccccll@{}}
\hline
UT date & MJD & SN & Epoch$^a$ & $B$ & $V$ & $R$ & $I$ & Telescope & Seeing$^b$ \\
\hline
2008 Sep 24 & 54\,733.2 & SN~2007if       & 358.2 & $24.89\pm0.65^c$ & $24.79\pm0.41$ & $25.14\pm0.65^c$ & $24.36\pm0.65^c$ & VLT  & 1.0 \\
2008 Sep 30 & 54\,739.8 & SNF &  54.8 & $20.72\pm0.16$   & $20.23\pm0.20$ & $20.01\pm0.10$   &                    & CAHA & 1.6 \\
2009 Feb 20 & 54\,882.2 & SNF & 187.3 & $23.35\pm0.30$   & $23.16\pm0.20$ & $23.17\pm0.31$   &                    & CAHA & 1.2 \\
2009 Apr 26 & 54\,947.4 & SNF & 248.0 & $24.18\pm0.17$   & $23.92\pm0.14$ & $24.76\pm0.12$   & $24.02\pm0.17$   & VLT  & 0.7 \\
2009 May 22 & 54\,973.2 & SNF & 272.0 & $24.40\pm0.14$   & $24.28\pm0.12$ & $25.25\pm0.22$   & $24.69\pm0.21$   & VLT  & 0.5 \\
\hline
\end{tabular}
\\[1.5ex]
\flushleft
$^a$~Phase in rest-frame days with respect to $B$-band maximum [MJD $= 54\,348.4$ 
     for SN~2007if \citep{scalzo2010a} and MJD $= 54\,680.9$ for SNF20080723-012 
     \citep{scalzo2012a}].\quad
$^b$~Stellar FWHM (arcsec).\quad
$^c$~Magnitudes obtained from an integration of the redshift-corrected 
     and flux-calibrated spectrum with Bessell $BRI$ filters.\\
\end{center}
\end{footnotesize}
\end{table*}

Since the method described above is very sensitive to errors related to the 
measurement or calibration of either the `SN + host' or host magnitudes, it 
provided questionable results in bands where the SN contribution was small 
compared to that of the host. We therefore decided to calibrate only the 
$V$-band SN magnitude in this way, since this is the band where the strongest 
emission lines are located in nebular SN~Ia spectra at $z = 0.074$, and where 
therefore the contrast between the SN and the background is best. Our $B$, $R$ 
and $I$ magnitudes were instead synthesised from our background-subtracted, 
flux-calibrated (with respect to the $V$-band photometry) spectrum of SN~2007if 
taken during the same night.

\subsection{Photometric calibration; $S$- and $K$-correction}
\label{Photometric calibration; $S$- and $K$-correction}

Our single-epoch photometry of SN~2007if was obtained under photometric 
conditions, and the zero points were derived from a \citet{stetson2000a} 
standard field observed on the same night. For 
SNF20080723-012 we had four epochs of photometry, not all of them obtained 
in photometric conditions. Therefore, a sequence of stars in the SN field 
(indicated in Fig.~\ref{fig:chart}, top) was calibrated with respect to 
\citet{stetson2000a} standards observed during the photometric nights on 2009 
April 26 and May 22 (Table~\ref{std}). The SN magnitudes in all individual 
nights were then determined relative to this sequence of stars.

To correct for deviations of the instrumental filter responses from the 
standard Johnson--Cousins systems \citep{bessell1990a} and the non-negligible 
redshift of our targets, $S$- and $K$-corrections (Table~\ref{SKcorr}) were 
derived from our nebular spectra and applied to the SN photometry. Since our 
photometric observations of SNF20080723-012 started already 55\,d after the 
$B$-band maximum, but we had no spectra at that epoch, we instead used a 
spectrum of SN~2007if [actually a combination of the +51\,d, +62\,d and 
+67\,d spectra presented by \citet{scalzo2010a}] to calculate the $S$- and 
$K$-corrections. The $S$- and $K$-corrections for the 187\,d photometry of 
SNF20080723-012 were then derived by linear interpolation between the 55\,d 
and 248\,d values.

The final, $S$- and $K$-corrected magnitudes of SN~2007if and SNF20080723-012 
are given in Table~\ref{mags}, along with their associated photometric errors.

\section{Luminosity evolution}
\label{Luminosity evolution}

\begin{table*}
\caption{Basic properties of proposed super-Chandrasekhar SNe~Ia.} 
\label{properties}
\begin{footnotesize}
\begin{center}
\begin{tabular}{@{}lcccc@{}}
\hline
                             &     SN~2006gz      &     SN~2007if      &  SNF20080723-012   &     SN~2009dc      \\
\hline
Redshift $z_\mathrm{hel}$    &     0.0237$^a$     &     0.0742$^b$     &     0.0745$^b$     &     0.0214$^c$     \\
Distance modulus $\mu$ (mag) & 35.03$\pm$0.06$^d$ & 37.45$\pm$0.05$^b$ & 37.46$\pm$0.05$^b$ & 34.86$\pm$0.08$^c$ \\
Colour excess \ebv\ (mag)    & 0.18$\pm$0.05$^a$  &    0.079$^{d,e}$   &     0.064$^e$      & 0.17$\pm$0.07$^c$  \\
\dm15                        & 0.69$\pm$0.04$^a$  & 0.71$\pm$0.06$^d$  & 0.93$\pm$0.04$^f$  & 0.71$\pm$0.03$^c$  \\
\hline
\end{tabular}
\\[1.5ex]
\flushleft
$^a$~\citet{hicken2007a}.\quad
$^b$~Heliocentric redshift from \Ha\ emission in the host galaxy, distance modulus from $z_\mathrm{hel}$.\quad
$^c$~\citet{silverman2011a,taubenberger2011a}.\quad
$^d$~\citet{scalzo2010a}.\quad
$^e$~Galactic colour excess \citep{schlegel1998a}.\quad\
$^f$~\citet{scalzo2012a}.\\
\end{center}
\end{footnotesize}
\end{table*}

\begin{figure*}
   \centering
   \includegraphics{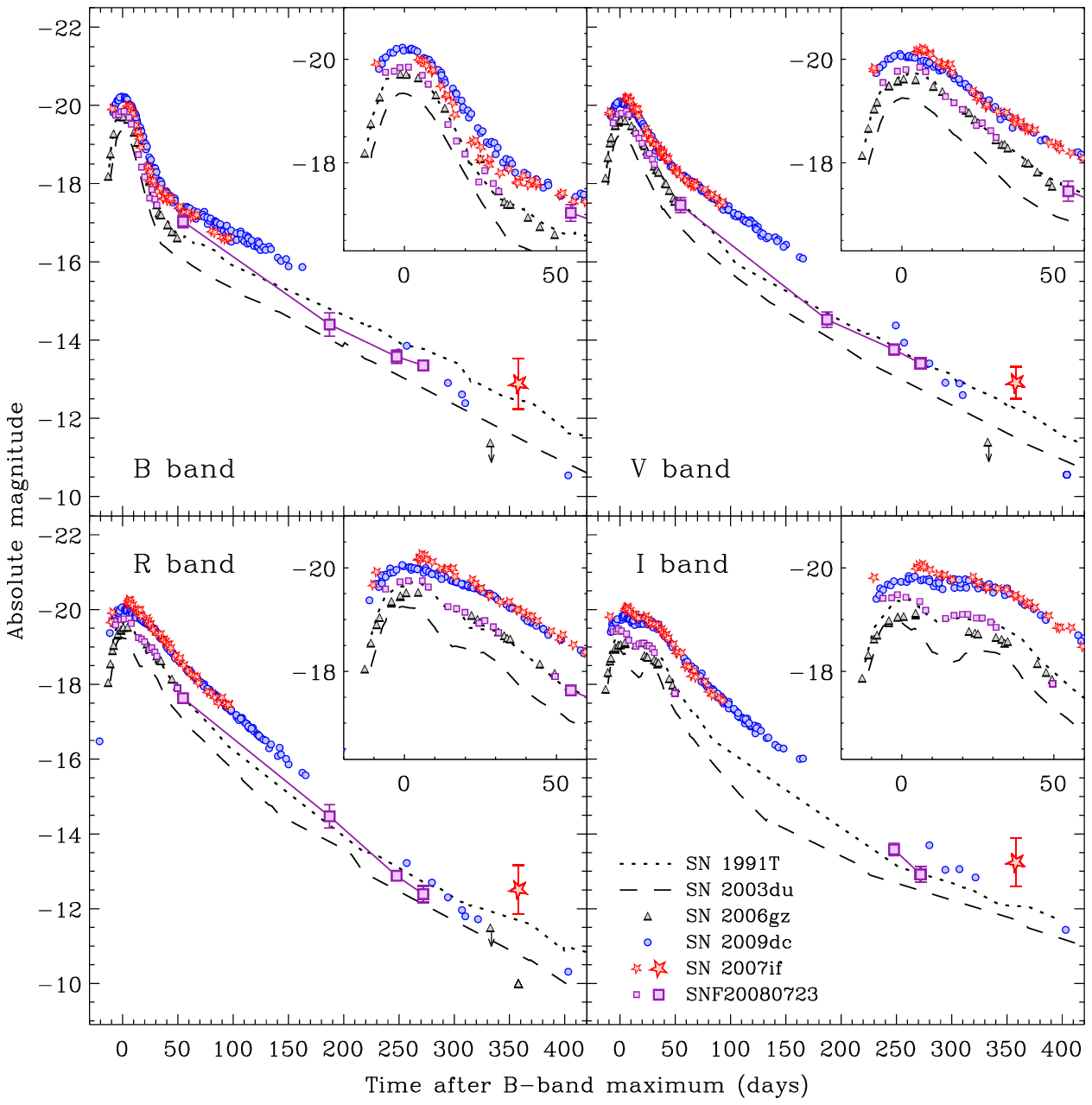}
   \caption{Absolute $BV\!RI$ light curves of the proposed super-Chandrasekhar 
   SNe~2006gz \citep{hicken2007a,maeda2009a}, 2007if \citep{scalzo2010a}, 
   2009dc \citep{silverman2011a,taubenberger2011a} and SNF20080723-012 
   \citep{scalzo2012a}, the `classical' luminous 
   SN~Ia 1991T \citep{lira1998a,altavilla2004a} and the normal SN~Ia 2003du 
   \citep{stanishev2007b}. Large symbols with error bars are data newly presented 
   in this work (Table~\ref{mags}). The phase is given in rest-frame days with 
   respect to $B$-band maximum.}
   \label{fig:LC}
\end{figure*}

Absolute $BV\!RI$ light curves of the candidate super-Chandrasekhar 
SNe~2006gz, 2007if, 2009dc and SNF20080723-012 are presented in 
Fig.~\ref{fig:LC}, and compared to those of SNe~2003du and 1991T as 
prototypes of a normal and a luminous SN~Ia, respectively. The absolute 
magnitudes shown in that figure were computed using the distance moduli 
and colour excesses reported in Table~\ref{properties}.

\begin{figure}
   \centering
   \includegraphics[width=8.4cm]{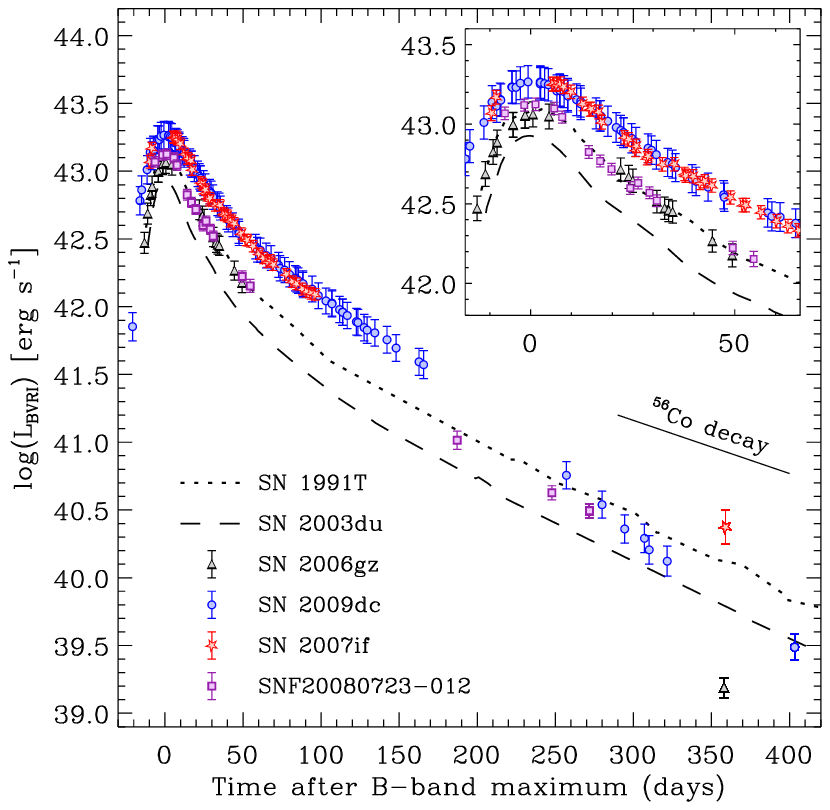}
   \caption{$BV\!RI$-integrated pseudo-bolometric light curves of the same 
   SNe as in Fig.~\ref{fig:LC}. The peak phase is enlarged in the inset. 
   Epochs are given in rest-frame days after $B$-band maximum.}
   \label{fig:bolo}
\end{figure}

It can readily be seen from the panels of Fig.~\ref{fig:LC} that for 
all SNe the late-phase $B$-, $V$-, $R$- and $I$-band light curves are 
quite similar to one another, so that the important trends are preserved 
when moving on to bolometric light curves. We therefore limit ourselves 
to discussing in detail the pseudo-bolometric light curves of all the 
objects, which are presented in Fig.~\ref{fig:bolo}. They
were obtained by transforming the absolute magnitudes into monochromatic 
luminosities at the effective wavelengths of the filters, interpolating 
the spectral energy distribution linearly and integrating over wavelength. 
Zero flux was assumed at the integration boundaries (i.e. the blue edge 
of the $B$ band and the red edge of the $I$ band). The resulting $BV\!RI$ 
bolometric light curves are expected to account for $\sim$\,50 -- 75 per 
cent of the total bolometric flux at maximum light when the IR contributes 
$\sim$\,10 per cent in SN~2009dc \citep{taubenberger2011a} and 
$\sim$\,10--15 per cent in normal SNe~Ia (Mezrag et al. in prep.), and the 
UV contribution blueward of the $U$ band should have dropped below the 
10 per-cent level \citep{silverman2011a}. At late phases, we may miss a 
significant amount of IR flux, but lack of late-time IR photometry of 
superluminous SNe~Ia prevents a more quantitative assessment.

Focusing on the early light curves (inset in Fig.~\ref{fig:bolo}), 
SN~2003du has the narrowest and least luminous peak 
($L_{BV\!RI}^\mathrm{\,max}\sim0.8\times10^{43}$ erg\,s$^{-1}$) in our 
sample of objects, in line with the expectations for a normal SN~Ia. 
SN~1991T is clearly more luminous and more slowly declining than SN~2003du, 
and interestingly quite similar to the proposed super-Chandrasekhar 
SNe~2006gz and SNF20080723-012 \citep[see also][]{scalzo2012a}. These 
three SNe reach peak luminosities of 
$L_{BV\!RI}^\mathrm{\,max}\sim1.2\times10^{43}$ erg\,s$^{-1}$.
Finally, SNe~2007if and 2009dc outshine all the other SNe, being more 
than twice as luminous as SN~2003du at peak 
($L_{BV\!RI}^\mathrm{\,max}\sim1.9\times10^{43}$ erg\,s$^{-1}$). 
The early-time pseudo-bolometric light curves of these two objects are 
essentially identical.

During the early nebular phase, between $\sim$\,100 and 200\,d after 
maximum, all SNe evolve as expected from their behaviour around peak 
brightness. SNe~2007if and 2009dc, which feature the brightest peak, 
are also most luminous at these phases. SNF20080723-012 and SN~1991T 
lie between normal SNe~Ia and SNe~2007if and 2009dc. The light-curve 
decline of all these SNe is quite similar, a bit faster in normal 
SNe~Ia and a bit slower in the superluminous SNe, but always somewhat 
faster than the decay rate of \Cofs, which is expected in the case of 
increasing $\gamma$-ray losses. No data are available for SN~2006gz 
during the early nebular phase. 

The simple hierarchical picture that seems to emerge from studying the 
luminosity evolution out to $\sim$\,200\,d, however, does not easily fit 
the data thereafter. SNe~2003du, 1991T, 2007if and SNF20080723-012 still 
decline with a nearly constant slope between 200 and 400\,d after maximum. 
Among these SNe the luminosity differences observed at earlier times are 
approximately preserved, though SNF20080723-012 now appears a little 
fainter than SN~1991T. SN~2006gz, on the other hand, fades dramatically 
some time between the peak phase and the nebular phase \citep{maeda2009a}, 
and a year after the explosion it is a factor of $\sim$\,4 less luminous 
than the normal SN~Ia 2003du and almost an order of magnitude less 
luminous than SN~1991T. SN~2009dc seems to share the destiny of SN~2006gz, 
though to a lesser extent. Around 200\,d after maximum its light curve 
starts to decline more rapidly than before 
\citep{silverman2011a,taubenberger2011a}, and, while not fading as 
rapidly as SN~2006gz, at 400\,d after maximum it is no longer more 
luminous than SN~2003du.

\section{Spectroscopic comparison}
\label{Spectroscopic comparison}

\begin{figure*}
   \centering
   \includegraphics[width=17.6cm]{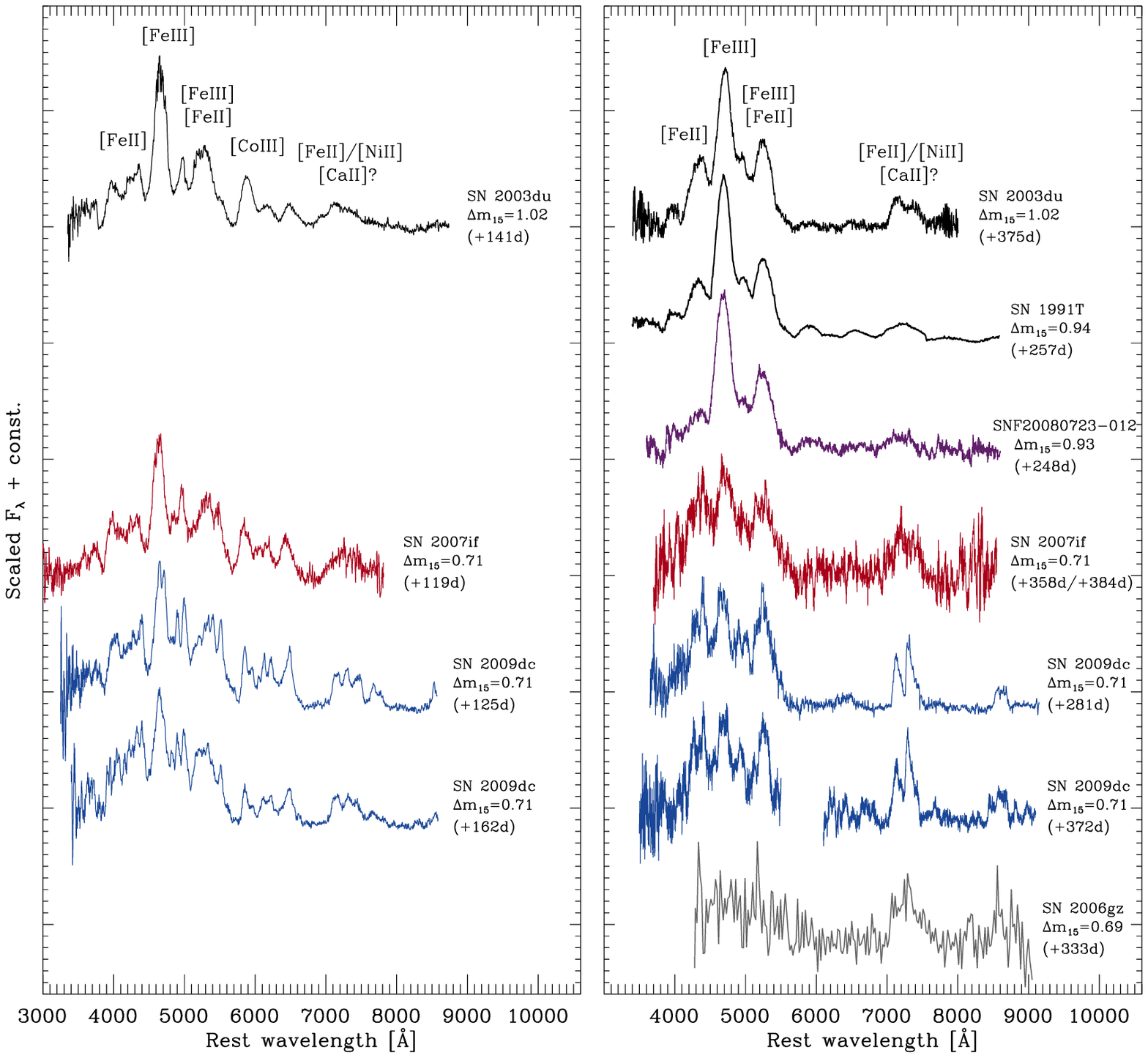}
   \caption{Nebular spectra of SNe~2007if (\citealt{blondin2012a}; this work), 
   2009dc (\citealt{taubenberger2011a,silverman2011a}; this work) and 
   SNF20080723-012 (this work; cf. 
   Table~\ref{spectra}), compared to those of the proposed super-Chandrasekhar 
   SN~2006gz \citep{maeda2009a}, the luminous SN~Ia 1991T \citep{gomez1998a} and 
   the normal SN~Ia 2003du \citep{stanishev2007b}. Phases are given in rest-frame 
   days with respect to $B$-band maximum. 
   {\bf Left panel:} Spectra $\sim$\,120--160\,d after maximum. For presentation 
   purposes, the spectrum of SNe~2007if and the $+162$\,d spectrum of SN~2009dc 
   have been smoothed by 500 and 2500 \kms, respectively.
   {\bf Right panel:} Spectra $\sim$\,250--400\,d after maximum. The two spectra 
   of SN~2009dc and the spectra of SN~2007if and SNF20080723-012 have been 
   smoothed by 300, 1300, 1000 and 1000 \kms, respectively. The gap in the 
   $+372$\,d spectrum of SN~2009dc is the region between the XShooter UVB and VIS 
   channels, with very low signal in both.}
   \label{fig:spectra}
\end{figure*}

Nebular spectra of our sample of superluminous SNe~Ia, SN~1991T and SN~2003du 
are shown in Fig.~\ref{fig:spectra}. In general, the spectral evolution during 
the nebular phase is slow, but not negligible. Since the spectra in 
Fig.~\ref{fig:spectra} have been taken at very different epochs, evolutionary 
effects and intrinsic differences between SNe have to be disentangled. This is 
straightforward for those objects for which multi-epoch spectroscopy is available.

Concentrating first on the early nebular phase (left panel of 
Fig.~\ref{fig:spectra}), we see that the superluminous SNe~Ia 2007if and 2009dc 
are relatively similar to each other and also to the normal SN~Ia 2003du. The 
same [\FeIII], [\FeII] and [\CoIII] emission lines are detected in all three SNe 
\citep[see e.g.][for a detailed identification of emission lines in nebular 
SN~Ia spectra]{maeda2010c,mazzali2011a}. 
Differences are evident in the width of spectral features, with many lines being 
resolved into double or multiple peaks in SN~2009dc that are just blended into 
a single broad feature in SN~2003du. Also the ionisation state is noticeably 
different, with weaker [\FeIII] and more prominent [\FeII] lines in SN~2009dc 
compared to SN~2003du. In both respects, SN~2007if takes an intermediate 
position between SNe~2003du and 2009dc, but in terms of ionisation closer to 
SN~2009dc. 

During the later nebular phase (right panel of Fig.~\ref{fig:spectra}) the 
observed differences become more pronounced. In normal SNe~Ia such as 
SN~2003du, the most important evolution between four months and one year after 
the explosion is the fading of Co emission lines due to the radioactive decay 
of \Cofs\ into \Fefs\ with $t_{1/2}(^{56}\mathrm{Co}) = 77.2$\,d 
\citep{kuchner1994a}. The same is also observed in the superluminous SNe~2007if 
and 2009dc, but here a lower ejecta ionisation state is seen as an additional 
effect. The hallmark feature of nebular SN~Ia spectra, the prominent [\FeIII] 
blend at $\sim$\,4700\,\AA\ \citep{axelrod1980a,spyromilio1992a,mazzali1998a}, 
is merely a stump in SNe~2007if and 2009dc. Instead, most of the emission 
blueward of 5500\,\AA\ probably originates from [\FeII] transitions. 

In normal SNe~Ia the features between 7000 and 7500\,\AA\ are attributed to 
forbidden transitions of iron-group 
elements (IGEs), notably [\FeII] $\lambda7155$ and [\NiII] $\lambda7378$, which 
often form a double-peaked structure as observed in SN~2003du 
\citep{maeda2010c,tanaka2011a}. 
In the nebular spectrum of SN~1991T, otherwise very similar to that of SN~2003du, 
there is only a single, rounded, broad peak from 7000 to 7500\,\AA, which may be 
the effect of a distribution of IGEs out to higher velocities in SN~1991T 
\citep{mazzali1995a}. In SNe~2007if and 2009dc the emission in that region is 
stronger than in the previously mentioned objects. The SN~2007if spectrum is 
too noisy to study the line profiles in detail, but in SN~2009dc there are two 
pronounced, sharp emission peaks. The blue peak can be explained 
by [\FeII] $\lambda7155$, whereas the redder at $\sim$\,7305\,\AA\ is not at the 
right position for [\NiII] $\lambda7378$ emission. We instead suggest a 
significant contribution of [\CaII] $\lambda\lambda7291,7324$. These lines are 
typically weak or absent in normal SNe~Ia, but have been identified in peculiar 
SNe~Ia, in particular in subluminous, 91bg-like events \citep{mazzali1997a,
mazzali2012a}. The fact that in SN~2009dc the 7000 to 7500\,\AA\ region shows 
an unusual triple-peaked structure at phases around day 150 (Fig.~\ref{fig:spectra}, 
left panel) suggests that [\CaII] emission might be present already at those 
epochs and strengthen with time.

The nebular spectrum of SN~2006gz shares similarities with those of SNe~2007if 
and 2009dc, especially in the strength of the 7000 to 7500\,\AA\ emission. However, 
the flux in the blue part of the spectrum is strongly suppressed, and individual 
features cannot be identified as a consequence of the low signal-to-noise ratio. 
Note that the SN~2006gz spectrum as presented here and in \citet{maeda2009a} 
is binned over 16\,px\,/\,22\,\AA.
The SNF20080723-012 spectrum, on the other hand, is very similar to the spectrum 
of SN~1991T, including the broad single-peaked emission between 7000 and 7500\,\AA. 
It thus deviates strongly from the spectra of the other proposed super-Chandrasekhar 
SNe~Ia. Taking into account the light-curve properties discussed in the previous 
section, one may therefore speculate that SNF20080723-012 might be a 91T-like object 
rather than a true member of the class of `super-Chandrasekhar' SNe~Ia, as already 
discussed by \citet{scalzo2012a}.

\begin{figure}
   \centering
   \includegraphics[width=8.4cm]{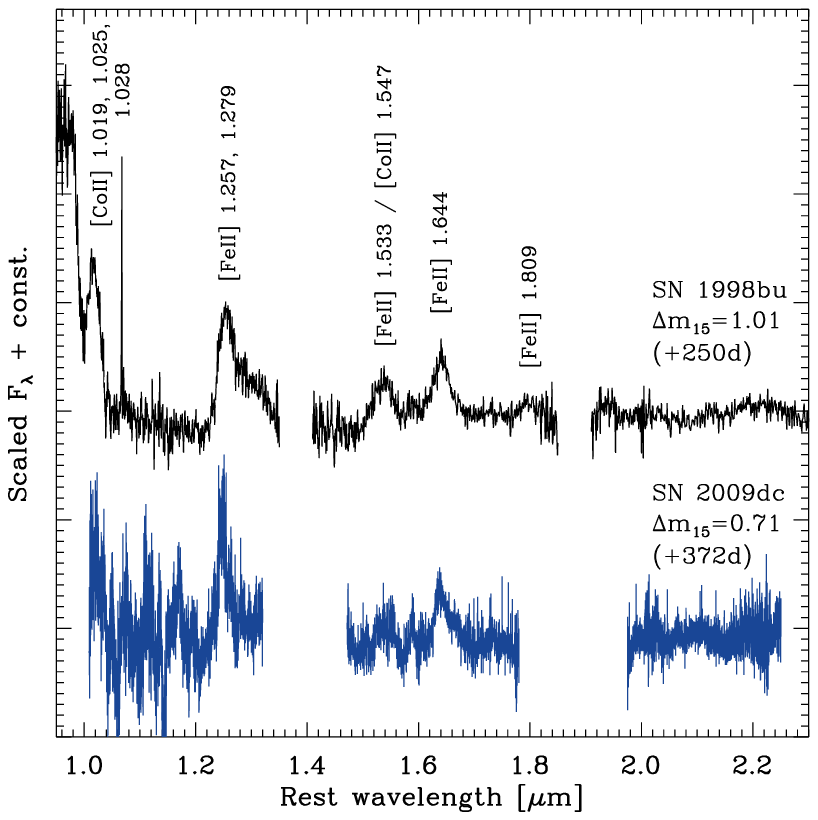}
   \caption{Nebular NIR spectrum of SN~2009dc, compared to the normal SN~Ia 1998bu 
   \citep{spyromilio2004a}. Phases are given in rest-frame days with respect to 
   $B$-band maximum. The spectrum of SN~2009dc has been smoothed by 2000 \kms, and 
   regions with strong telluric features are omitted.}
   \label{fig:NIR}
\end{figure}

At NIR wavelengths (Fig.~\ref{fig:NIR}) only a few features can be safely identified 
in the spectrum of SN~2009dc as a consequence of the low signal-to-noise ratio. 
The detected [\FeII] 1.257, 1.279\,$\mu$m, [\FeII] 1.533\,$\mu$m\,/\,[\CoII] 
1.547\,$\mu$m and [\FeII] 1.644\,$\mu$m emission lines are characteristic of nebular 
NIR spectra of SNe~Ia, and prominent also in the normal SN~Ia 1998bu \citep{spyromilio2004a} 
which is included in Fig.~\ref{fig:NIR} for comparison. A contribution of [\SiI] 
1.646\,$\mu$m to the 1.64\,$\mu$m line \citep[e.g.][]{mazzali2011a} cannot be excluded 
in SN~2009dc, though the feature is not stronger than in SN~1998bu where it was 
explained by [\FeII] alone \citep{spyromilio2004a}.
An additional emission feature is identified in SN~2009dc at $\sim$\,1.17\,$\mu$m, 
which has no discernible counterpart in SN~1998bu. It is located in a region devoid 
of strong telluric absorptions, and is similarly pronounced as other lines discussed 
before. Hence, we tend to believe that it is not a mere reduction artefact, but 
cannot provide a conclusive identification without detailed modelling.

\section{Discussion}
\label{Discussion}

\subsection{Ionisation state in the nebular phase}
\label{Ionisation state in the nebular phase}

\begin{figure}
   \centering
   \includegraphics[width=8.4cm]{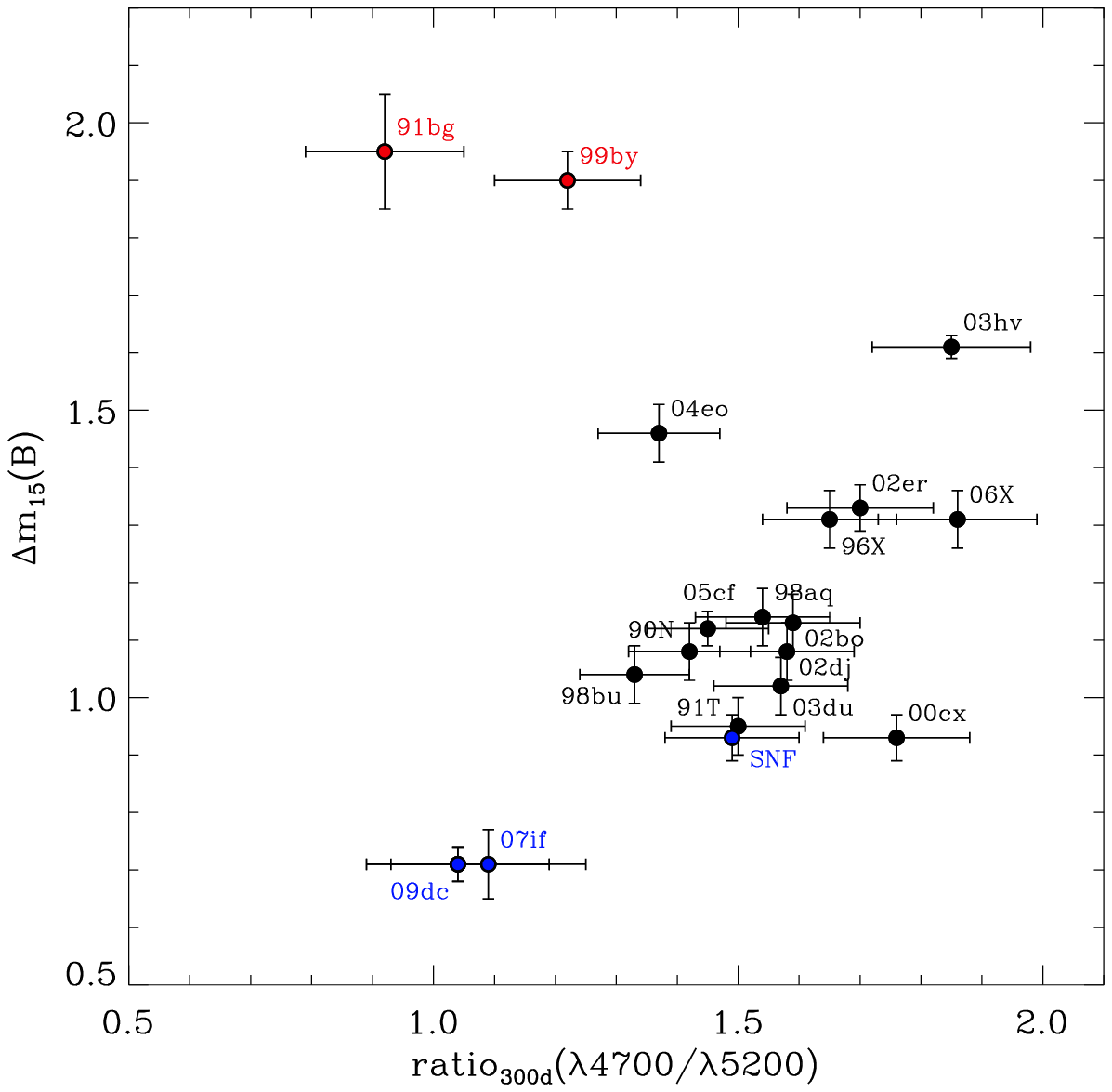}
   \caption{Relation between \dm15\ and the flux ratio of the [\FeIII]\,/\,[\FeII] 
   emission lines near 4700 and 5200\,\AA, determined in the nebular spectra closest 
   to day 300 after maximum. The line fluxes have been measured from zero 
   intensity through a multiple-Gaussian deblending of the 4500--5500\,\AA\ region 
   as described by \citet{mazzali1998a}. Red symbols stand for subluminous SNe~Ia, 
   blue symbols for proposed super-Chandrasekhar events, and black symbols for 
   normal or 91T-like SNe~Ia.}
   \label{fig:ratio}
\end{figure}

The ionisation state of SN~Ia ejecta during the nebular phase can be assessed 
by studying the flux ratio of the emission blends close to 4700 and 5200\,\AA. As 
discussed in detail e.g. by \citet{mazzali2011a}, the 4700\,\AA\ blend is clearly 
dominated by [\FeIII] emission, whereas the 5200\,\AA\ blend has a significant 
contribution of [\FeII] next to [\FeIII], making the ratio a sensitive ionisation 
indicator. In order to reproduce the large observed $\lambda4700/\lambda5200$ 
ratio in the normal or slightly subluminous SN~2003hv with synthetic spectra, 
\citet{mazzali2011a} had to assume a reduced central density compared to that 
of the one-dimensional Chandrasekhar-mass explosion model W7 \citep{nomoto1984a}, 
leading these authors to speculate about a sub-Chandrasekhar-mass or merger origin 
of SN~2003hv.
In an analogous way, the low ionisation found in SNe~2007if and 2009dc might be 
indicative of high central ejecta densities, leading to enhanced recombination. 
This should be a direct consequence of the low ejecta expansion velocities 
determined from both early- and late-time spectra (\citealt{scalzo2010a,
yamanaka2009a,silverman2011a,taubenberger2011a}; see also 
Section~\ref{Spectroscopic comparison}).

In Fig.~\ref{fig:ratio} the flux ratio of the emission lines near 4700 and 
5200\,\AA\ is investigated as a function of \dm15 [which in normal SNe~Ia may be 
taken as a proxy for the peak luminosity; see \citet{phillips1993a}]. Normal and 
91T-like SNe (including SNF20080723-012) show flux ratios between 1.3 and 1.9, 
SNe~2007if and 2009dc between 1.0 and 1.1. The only other SNe with similarly low 
flux ratios (and hence ionisation) are subluminous 91bg-like SNe. Their nebular 
spectra, however, show a complex structure with very narrow [\FeIII] lines 
superimposed on broad [\FeII] emission \citep{mazzali1997a,mazzali2012a}.

The inferred low ionisation is also consistent with the likely detection of 
[\CaII] $\lambda\lambda7291,7324$. The first and second ionisation potentials of 
Ca are lower than those of Fe. Accordingly, in regions with significant \FeIII\ 
content, Ca should be doubly ionised to almost 100 per cent, which is probably 
why [\CaII] lines are absent from the nebular spectra of normal SNe~Ia. The
lower ionisation in superluminous and 91bg-like SNe~Ia, however, should favour 
\CaII\ as the dominant ionisation state. Indeed, [\CaII] emission is observed
in exactly those objects.

\subsection{Late-time dust formation?}
\label{Late-time dust formation?}

\begin{figure*}
   \centering
   \includegraphics[width=16.5cm]{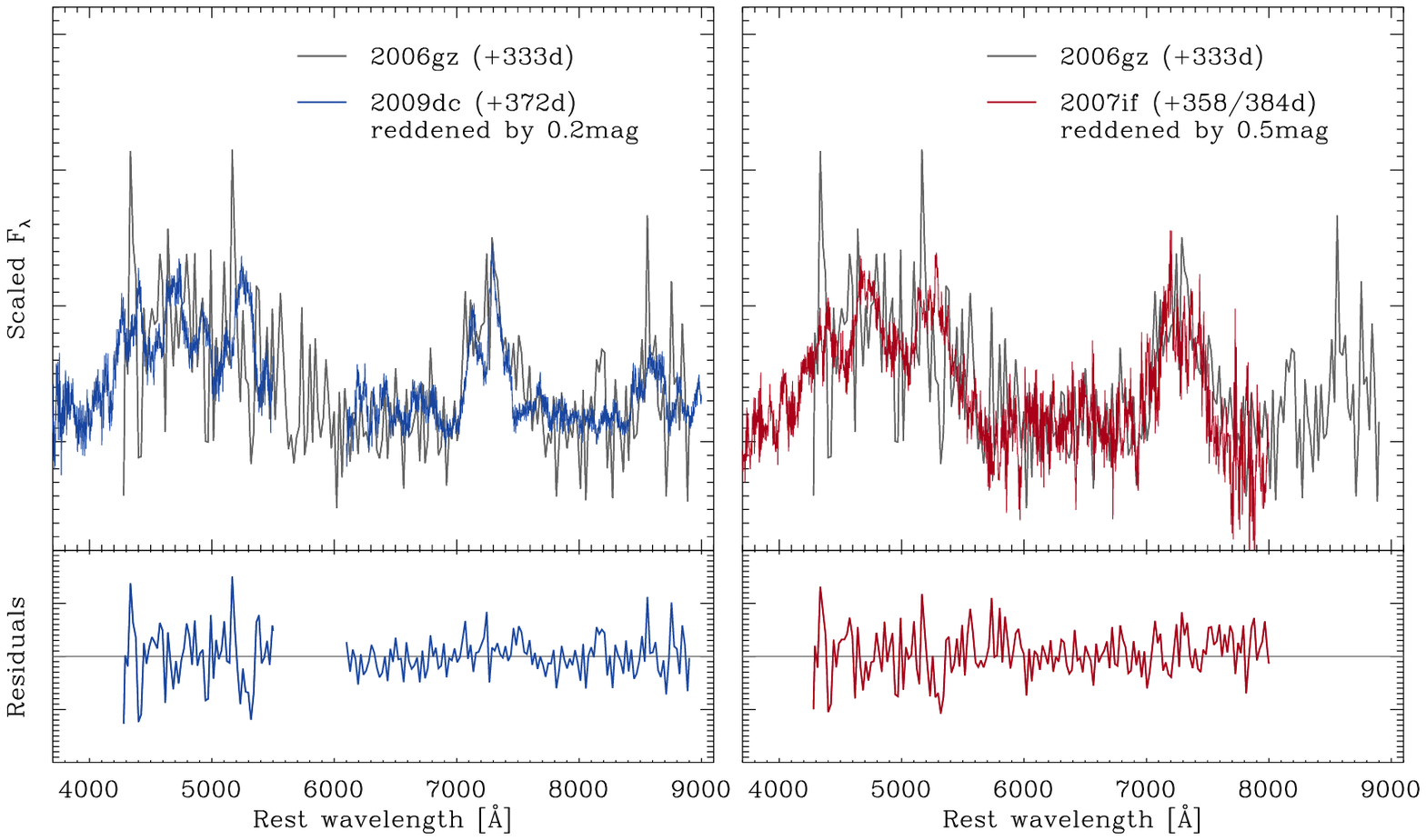}
   \caption{Comparison of the late-time spectrum of SN~2006gz with that of 
   SN~2009dc, artificially reddened by \ebv\ = 0.2 mag (upper left panel), 
   and that of SN~2007if, artificially reddened by \ebv\ = 0.5 mag (upper 
   right panel; $R_V=3.1$ assumed in both cases). The residuals 
   after subtracting the spectra (lower panels) hardly show any differences 
   exceeding the noise level.}
\label{fig:06gz_09dc_07if}
\end{figure*}

In Section~\ref{Luminosity evolution} we discussed the diverse light-curve 
decline of superluminous SNe~Ia at late phases. Once on the radioactive tail 
$\sim$\,50\,d after maximum light, SN~2007if fades at a nearly constant rate 
for more than 300\,d. SN~2009dc shows a steeper decline after $\sim$\,200\,d, 
and is about a factor 3 less luminous than SN~2007if after one year, though 
it had the same peak luminosity (Fig.~\ref{fig:bolo}). SN~2006gz fades even 
more rapidly, being about a factor 15 less luminous than SN~2007if one year 
after maximum, though the difference at peak was merely a factor 1.6. 

Following the reasoning of \citet{taubenberger2011a}, the accelerated decline 
observed in the $BV\!RI$-bolometric light curve of SN~2009dc (and probably 
also SN~2006gz) at those late phases is unlikely to be a true bolometric effect. 
As long as the light-curve tail is powered by radioactive decay and positrons 
are fully trapped, the decline may slow down (when a longer-lived radioactive 
isotope starts to dominate the energy deposition), but not accelerate [cf. 
\citet{ruiz-lapuente1998a} for the case of incomplete positron trapping]. 
Instead, the observed luminosity drop is probably the outcome of flux 
redistribution into regimes that are not observed, most likely the IR. This 
could be accomplished by an early IR catastrophe or by dust formation. 

Though never observed in SNe~Ia to date, an IR catastrophe \citep{axelrod1980a,
fransson1996a} is an inevitable consequence of the decreasing energy 
deposition by radioactive decay and the expansion of the SN ejecta. Below a 
critical temperature $T_\mathrm{c}$ the upper levels of forbidden transitions 
in the optical and NIR can no longer be populated. The cooling is henceforth 
dominated by ground-state fine-structure transitions of Fe in the mid- and 
far-IR, accompanied by a rapid temperature decrease of the ejecta. An IR 
catastrophe is favoured by low densities (as a consequence of the density 
dependence of $T_\mathrm{c}$). In the high-density ejecta of SN~2009dc an 
early onset of the IR catastrophe is therefore not expected. Moreover, even 
one year after maximum, when the $BV\!RI$-bolometric luminosity of SN~2009dc 
is already significantly below that of SN~2007if, the spectra are still 
similar, showing prominent [\FeII] emission throughout the optical and NIR 
wavelength range.

Dust formation, on the other hand, is usually associated with core-collapse 
SNe and has rarely been discussed in the context of thermonuclear SNe 
\citep{nozawa2011a}. In superluminous SNe~Ia, however, it seems to be
compatible with observations. Early-time spectra of SNe~2006gz and 2009dc 
show \CII\ lines with a strength unprecedented in thermonuclear SNe. This 
suggests that at least a moderate amount of carbon is present in the ejecta 
[\citet{hachinger2012a} reproduced the spectral time series of SN~2009dc 
with carbon mass fractions between one and ten per cent].
The carbon may give rise to the formation of graphite dust, but can also 
pave the way for dust formation in general through prior CO molecule formation, 
which opens an efficient cooling channel, the emission in molecular bands. 
Dust formation should also be promoted by the comparatively high densities 
that the slowly expanding ejecta of superluminous SNe retain at late phases 
\citep{nozawa2011a}. The luminosity drop in SN~2009dc comes along with an 
evolution towards redder colours (Fig.~\ref{fig:LC}), again consistent with 
dust formation. Interestingly, even the seemingly unique nebular spectrum of 
SN~2006gz with its apparent lack of features in the blue could find a 
natural explanation within this scenario. This is shown in 
Fig.~\ref{fig:06gz_09dc_07if}, where the late-time spectra of SNe~2007if and 
2009dc have been artificially reddened to simulate the effect of more 
pronounced dust formation in SN~2006gz. Nothing can be said about individual 
features blueward of $\sim$\,6000\,\AA\ since the spectrum of SN~2006gz is 
too noisy in that region, but from the overall spectral shape it is plausible 
that the late-time spectra of SNe~2006gz, 2007if and 2009dc are all 
intrinsically similar, and differ just by the amount of reddening caused by 
newly formed dust in the ejecta.

Note that changes in the profiles of nebular emission lines, usually another 
signature of dust formation within the ejecta, are neither observed nor 
expected in SN~2009dc. 
If dust formation occurs, it would be most effective in the carbon- and 
silicon-rich zones, which, according to the velocity evolution of the \CII\ 
$\lambda6580$ and \SiII\ $\lambda6355$ lines presented by 
\citet{taubenberger2011a}, should be located outside $\sim$\,6000 \kms. 
The [\FeII] emission in the nebular spectrum of SN~2009dc, however, comes 
from below 6000 \kms; emission above that velocity would result in the nebular 
lines being too broad. Accordingly, all parts of the nebular [\FeII] emission 
would be attenuated by the same amount.

\begin{figure*}
   \centering
   \includegraphics[width=17.6cm]{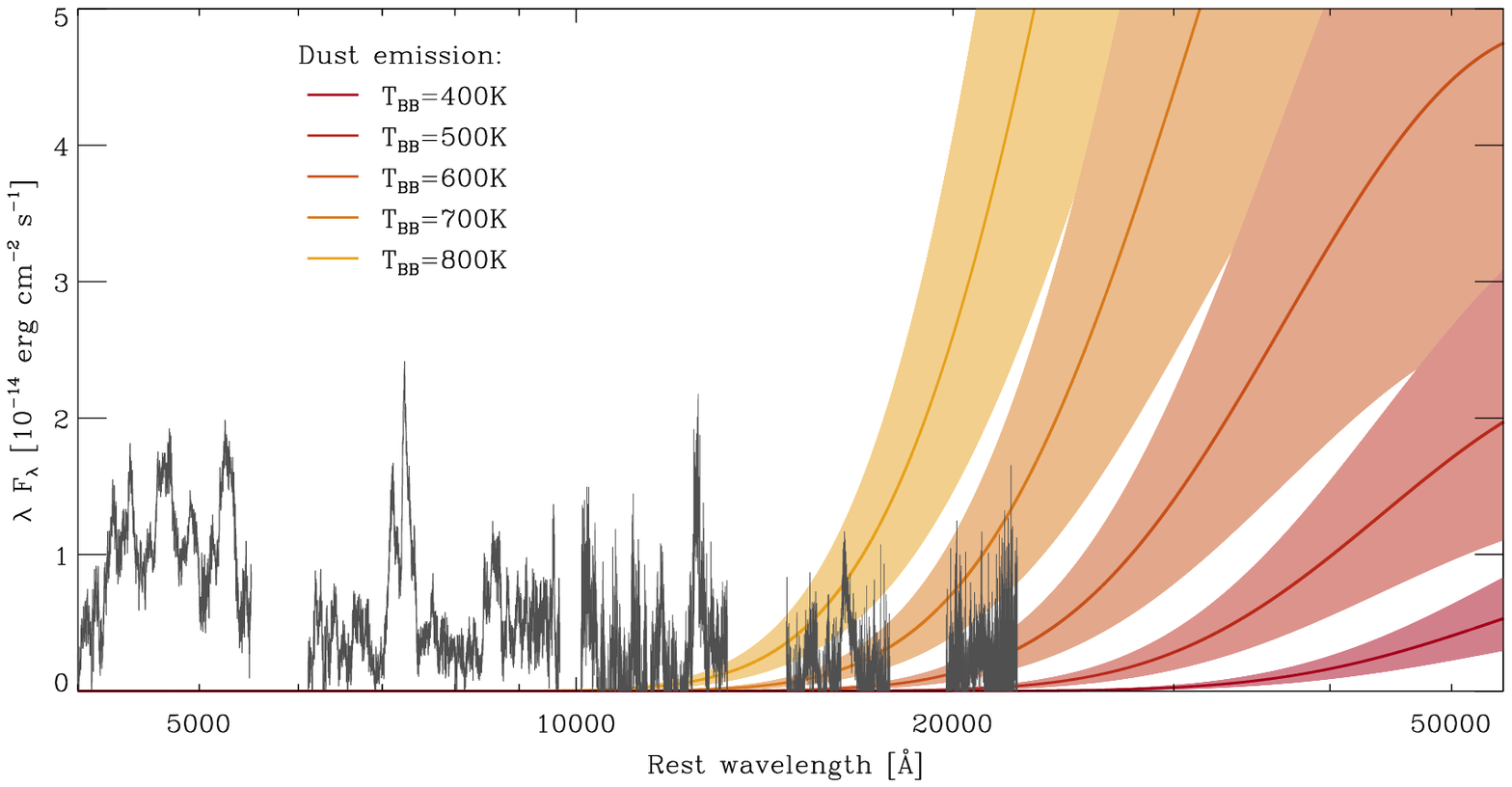}
   \caption{Nebular (+372\,d) spectrum of SN~2009dc, corrected for
   the Galactic and host-galaxy reddening of $E(B-V)=0.17$ mag. The solid lines 
   are blackbody spectra for temperatures $T_\mathrm{BB}$ between 400 and 800 K, 
   assuming emission from a thin dust shell at 8000 \kms. The lighter shaded 
   regions show the effect of varying the shell location between 6000 and 10\,000 
   \kms. It appears that a dust temperature of $\geq 700$\,K is excluded by a 
   comparison to the observed NIR spectrum of SN~2009dc.}
   \label{fig:dust}
\end{figure*}

Under the premise that dust formed in the ejecta of SN~2009dc, some of its 
properties can be estimated. To calculate the dust mass resulting in an extinction 
$A_V = 0.93$ mag [$E(B-V) = 0.3$ mag, $R_V = 3.1$] on day 372 after peak, we have 
assumed that the dust is located in an (infinitesimally) thin shell at radius $R$. 
For this geometry, the optical depth that corresponds to this value of $A_V$ 
(i.e. $\tau_V = 0.86$) can be expressed as
\begin{equation}
\label{eq:tau}
\tau_V = \frac{Q_V(a)\, \pi a^2 M_\mathrm{dust}}{4 \pi R^2\, m_\mathrm{grain}},
\end{equation}
where $a$ is the dust-grain radius, $m_\mathrm{grain}$ the dust-grain mass, 
$M_\mathrm{dust}$ the total dust mass, and $Q_V(a)$ the extinction efficiency
in the $V$ band. It follows that 
\begin{equation}
\label{eq:mdust}
M_\mathrm{dust} = \frac{16\pi}{3} R^2\, \frac{\rho_\mathrm{grain}\, a}{Q_V(a)}\, \tau_V,
\end{equation}
very similar to eq.~6 of \citet{lucy1989a}, but with a different prefactor 
reflecting the different geometry (a thin shell here versus a homogeneous 
sphere there). For small grains $Q_V(a)$ is approximately proportional to $a$ 
\citep{lucy1989a}, and hence the expression for $M_\mathrm{dust}$ becomes 
largely independent of $a$.
Assuming a dust-grain density of $\rho_\mathrm{grain} = 2.9$ g\,cm$^{-3}$ 
(appropriate for a mixture of graphite and silicates\footnote{When conditions are 
appropriate for dust formation in superluminous SNe~Ia, the presence of carbon and
also silicon may allow for the formation of carbonaceous dust as well as silicates 
(typical densities 2.2 and 3.5 g\,cm$^{-3}$, respectively; \citealt{weingartner2001a}). 
Given our ignorance of the exact dust composition we assume a 1:1 mixture in our 
calculation; the resulting uncertainty in the estimated dust mass is less than 
$\pm 25$ per cent.}), a grain size of 0.01 
$\mu$m, and adopting a $Q_V$ of 0.07 from fig.~4a of \citet{draine1984a}, we 
have evaluated Eq.~\ref{eq:mdust} for different radii of the dust shell, 
between 6000 and 10\,000 \kms\ in velocity space. The resulting dust mass 
ranges from $\sim$\,$1 \times 10^{-4}$ to $\sim$\,$4 \times 10^{-4}$ \Msun, 
similar to what has been reported for some core-collapse SNe in the literature 
\citep[e.g. for SN~2004et;][]{kotak2009a}.

The newly formed dust should also manifest in the emission of a thermal 
continuum reflecting the temperature of the dust grains. The emission would 
likely peak in the mid IR, but depending on the dust temperature the NIR 
regime may also be affected by the Wien tail of the blackbody spectrum.
To derive limits on the dust temperature in SN~2009dc, we have calculated 
blackbody curves for different temperatures $T_\mathrm{BB}$ and radii $R$ 
of 6000 to 10\,000 \kms\ in velocity space. The luminosity can be expressed as
\begin{equation}
\label{eq:lum}
L \approx 4 \pi^2 R^2\, B_\nu(T_\mathrm{BB}).
\end{equation}
In Fig.~\ref{fig:dust} the derived blackbody curves are overplotted on the 
optical-through-NIR spectrum of SN~2009dc taken 372\,d after maximum light.
The figure suggests that dust temperatures of 700\,K or more are excluded by 
the continuum level of the NIR spectrum, whereas $\leq 600$\,K may 
be acceptable. Such temperatures correspond to blackbody spectra peaking in the 
mid IR (at $\ga 4\ \mu$m). However, the uncertainties in this estimate are large: 
not only the radius of the dust shell is poorly constrained, also the flux 
calibration of the NIR spectrum of SN~2009dc is uncertain by a factor of $\sim$\,2 
(see Section~\ref{Observations and data reduction}).

In the end, mid-IR observations of superluminous SNe~Ia during the nebular 
phase will be the only way to prove or disprove dust formation on solid grounds. 
If dust forms, excess emission in the mid IR with a thermal spectral energy 
distribution should be detected.

\subsection{Bolometric light-curve models}
\label{Bolometric light-curve models}

\begin{figure}
   \centering
   \includegraphics[width=8.4cm]{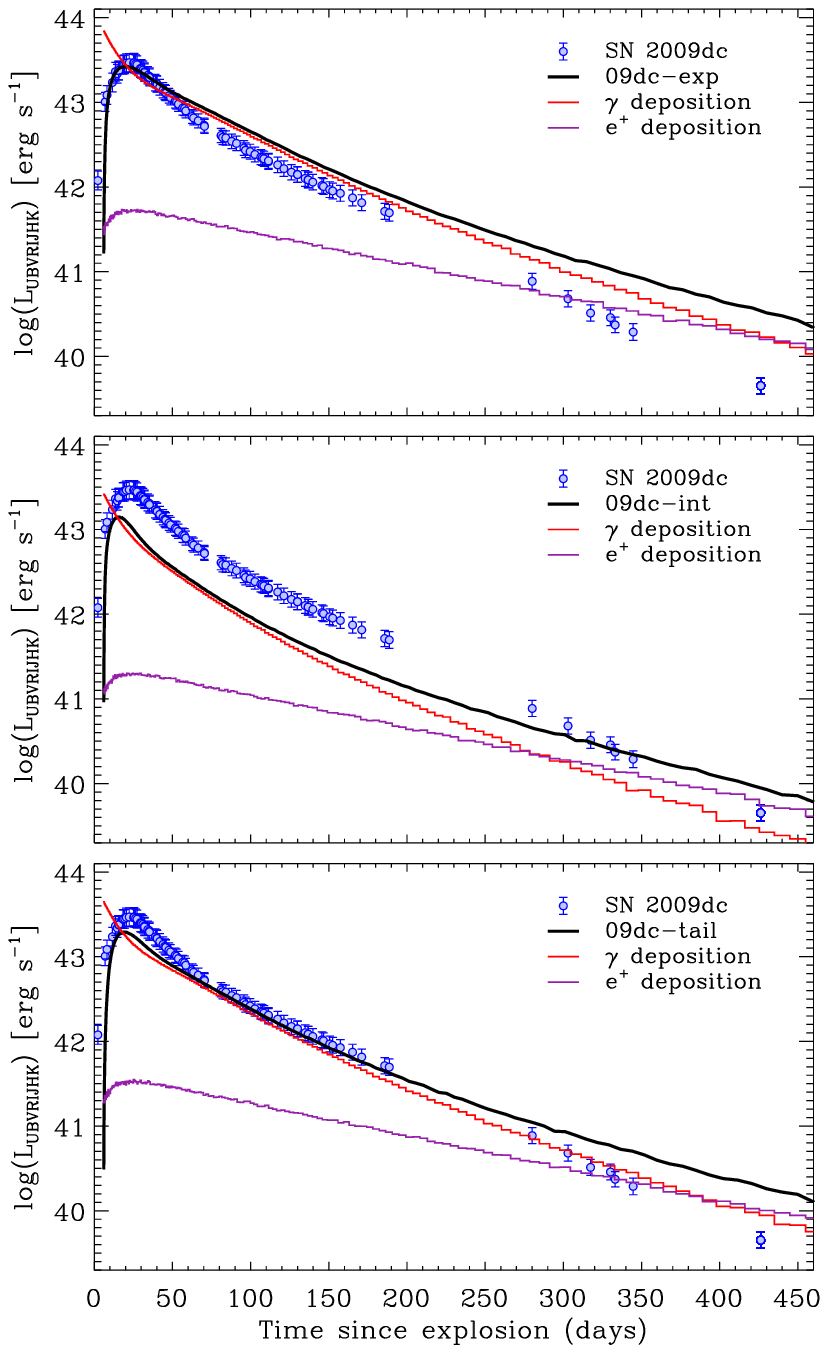}
   \caption{Synthetic bolometric light curves (black solid lines) for different 
   models compared to the observed $U\!BV\!RIJHK$-bolometric light curve of 
   SN~2009dc \citep[][blue data points, assuming a $B$-band rise time of 
   23\,d]{taubenberger2011a}. The contributions of 
   $\gamma$-rays (red solid lines) and positrons (purple solid lines) to the 
   synthetic bolometric light curves are shown individually.\newline {\bf Top panel:} The 
   09dc-exp model of \citet{hachinger2012a}.\newline {\bf Middle panel:} The 09dc-int model 
   of \citet{hachinger2012a}.\newline {\bf Bottom panel:} 09dc-tail ($\sim$\,2 \Msun\ 
   of ejecta, $\sim$\,1 \Msun\ of \Nifs).}
   \label{fig:models}
\end{figure}

The bolometric light curves presented in Section~\ref{Luminosity evolution} can 
serve as a benchmark for models proposed for superluminous SNe~Ia. To enable such 
a model--data comparison, we have computed synthetic bolometric light curves for 
a number of models using the Monte-Carlo radiative-transfer code {\sc artis} 
\citep{kromer2009a}. Full spectral calculations at late phases are not possible 
with {\sc artis}, since non-thermal processes and a detailed treatment of collisions 
and forbidden transitions are presently not implemented. 
However, for the late-phase bolometric light curve the deposition of $\gamma$-rays 
and positrons is the only relevant physical process. This is treated with 
sufficient accuracy in {\sc artis}. We have run our calculations with a simple grey 
optical opacity \citep{sim2007b}, which is a good approximation for the late 
bolometric light curve when time-dependent effects are negligible 
\citep{cappellaro1997a}. The rise time, the detailed shape of the light-curve peak 
and the peak luminosity, however, may be altered by this approximation.

For our light-curve calculations we have used the same models as in 
\citet{hachinger2012a}. These authors used early-time spectra of SN~2009dc to 
perform a tomographic study of the outer layers of the ejecta for three explosion 
scenarios. The first type of model investigated is the detonation of a rotating 
supermassive WD with 2 \Msun, represented by an AWD3det-based density structure 
\citep[][Fink et al. in prep.]{pfannes2010b}. The two alternative scenarios are a 
possible core-collapse origin of SN~2009dc, represented by an empirically derived 
exponential density structure with 3 \Msun\ (`09dc-exp'), and a model where part 
of the luminosity is assumed to be generated by CSM interaction, based on the W7 
\citep{nomoto1984a} density structure with 1.4 \Msun\ of ejecta but a rescaled 
kinetic energy (`09dc-int'). For the AWD3det-based model, \citet{hachinger2012a} 
were not able to obtain a consistent solution: the line velocities in the synthetic 
spectra were too high, and the inferred abundance structure was inconsistent with 
the kinetic energy of the underlying density profile. Accordingly, this model 
was disfavoured on the basis of the early spectral evolution. For the 09dc-exp and 
09dc-int models, on the other hand, more plausible solutions could be obtained. 

In their tomographic study, \citet{hachinger2012a} used the luminosity mostly as 
a free parameter. Synthetic bolometric light curves thus provide an additional 
constraint on the structure of the SN by constraining the total energy of the 
explosion. To compute bolometric light curves for the 
models of \citet{hachinger2012a}, an assumption had to be made for the inner core 
of the ejecta (below $\sim$\,4000 \kms) whose composition was not constrained by 
the tomography. Here we kept the mass fraction of stable iron the same as in the 
innermost shell studied by Hachinger et al., and filled the rest with \Nifs.

Fig.~\ref{fig:models} shows synthetic bolometric light curves of the 09dc-exp 
model (top panel) and the 09dc-int model (middle panel), compared to the 
$U\!BV\!RIJHK$ pseudo-bolometric luminosity evolution of SN~2009dc as presented 
by \citet{taubenberger2011a}. Clearly, the 09dc-exp model manages to reproduce 
the peak luminosity of SN~2009dc (which is a consequence of the large \Nifs\ 
mass of almost 1.6 \Msun), but it is also evident that the radioactive tail 
is much too bright in the model owing to the large total mass of 3 \Msun\ that 
leads to strong $\gamma$-ray trapping. This is further illustrated by the red and 
purple lines in Fig.~\ref{fig:models}, which give the individual contributions of 
energy deposited by $\gamma$-rays and positrons, respectively. The intersection thus 
marks the epoch where the $\gamma$-ray trapping has decreased to $\sim$\,3 per cent, 
which is approximately the fraction of \Cofs\ decay energy carried by positrons. 
In 09dc-exp this level is reached very late, $\sim$\,440\,d after the explosion. 
In contrast, the 09dc-int model fails to reach the observed peak luminosity of 
SN~2009dc by quite some margin, though in this case by purpose: the remainder is 
supposed to be generated by ejecta--CSM interaction, and not included in the 
synthetic bolometric light curve that shows only the contribution from radioactive 
decay. However, also the light-curve tail of 09dc-int seems much too dim at phases 
between 100 and 200\,d when the spectra with their lack of pseudo-continuum 
emission do not support significant ejecta--CSM interaction 
\citep{taubenberger2011a}. The model, as considered here, has only 
Chandrasekhar mass, i.e. it represents the exploded WD without any swept-up 
material \citep[cf.][]{hachinger2012a}. The low mass leads to inefficient 
$\gamma$-ray trapping (reaching 3 per cent already 280\,d after the explosion), 
insufficient to explain the observed luminosity between 100 and 200\,d.

\begin{figure}
   \centering
   \includegraphics[width=8.4cm]{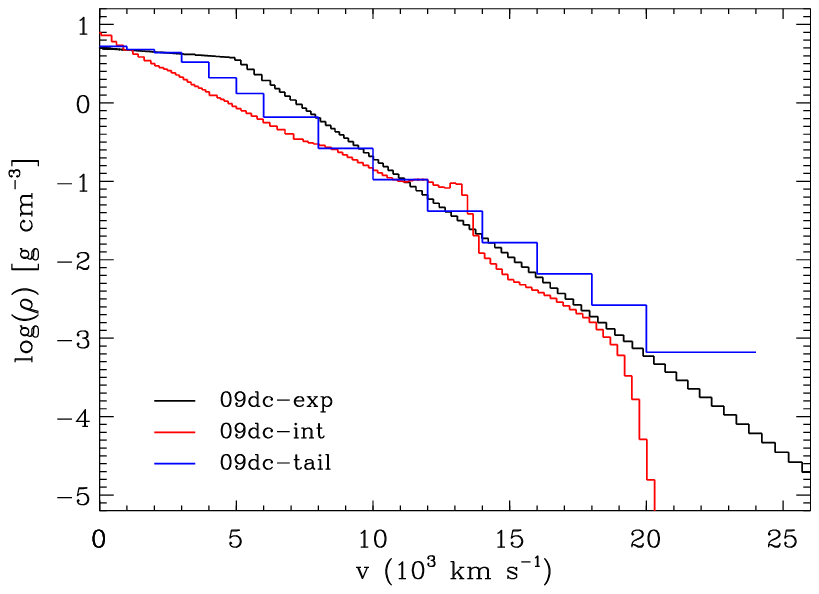}
   \caption{Density profiles of the models shown in Fig.~\ref{fig:models}, 
   evaluated at a reference time of 100\,s after the explosion.}
   \label{fig:rho}
\end{figure}

We now test whether the shortcomings of 09dc-int in fitting the light-curve tail 
before the break at $\sim$\,200\,d can be overcome by adding more mass. To this 
end, we have constructed a new model, `09dc-tail' (Table~\ref{09dc-blablub} and 
bottom panel of Fig.~\ref{fig:models}). Its density profile $\rho(v)$ is a moderately 
steep exponential for $v \geq 2500$ \kms, but flattens below that velocity. This 
results in a total ejecta mass of $\sim$\,2 \Msun, and a kinetic energy of 
$1.2\times10^{51}$ erg. The density profile of 09dc-tail is shown in 
Fig.~\ref{fig:rho}, together with those of 09dc-exp and 09dc-int. The \Nifs\ mass 
of 09dc-tail was adjusted to be 1 \Msun, more than in 09dc-int ($\sim$\,0.6 \Msun) 
but significantly less than in 09dc-exp ($\sim$\,1.6 \Msun). Thanks to the 
fairly steep density profile and the high ejecta mass the $\gamma$-ray opacity is 
large, resulting in 3 per cent $\gamma$-ray trapping at 390\,d after the explosion. 
The bolometric light curve of the model provides a very good fit to the data of 
SN~2009dc between 60 and 190\,d after the explosion. The model remains more 
luminous than SN~2009dc at later epochs, but as discussed in the previous section 
this may be due to a significant IR flux that was missed when the 
pseudo-bolometric light curve of SN~2009dc was constructed from the observed bands. 
As 09dc-int, the model does not reach the observed peak luminosity of SN~2009dc. 
It peaks at a flux 30--40 per cent too low.

\begin{table*}
\caption{
Density profile and element abundances of 09dc-tail. 
The total mass of the model, 2.06\,\Msun, is distributed to 
C~(0.24\,\Msun), O~(0.47\,\Msun), Si~(0.25\,\Msun), S~(0.05\,\Msun),
Ca~(0.01\,\Msun), stable~Fe~(0.05\,\Msun) and \Nifs\ (0.99\,\Msun).
} 
\label{09dc-blablub}
\begin{footnotesize}
\begin{center}
\begin{tabular}{@{}rrrccccccccc@{}}
\hline
Shell & $v_\mathrm{out}^a$ & $\log(\rho)^b$ & $M_\mathrm{cum}^{\,c}$ & $E_\mathrm{k,cum}^{\,d}$ &  X(C)  &  X(O)  &  X(Si) &  X(S)  &  X(Ca) & X(stable Fe) & X(\Nifs) \\
\hline
 1    &   1000             &   $0.72$       &    0.011               &   $6.6\times10^{-5}$     & 0.000  & 0.000  & 0.000  & 0.000  & 0.000  & 0.050        & 0.950    \\
 2    &   2000             &   $0.68$       &    0.082               &   $1.9\times10^{-3}$     & 0.000  & 0.000  & 0.000  & 0.000  & 0.001  & 0.049        & 0.950    \\
 3    &   3000             &   $0.64$       &    0.256               &      0.014               & 0.000  & 0.000  & 0.000  & 0.000  & 0.002  & 0.048        & 0.950    \\
 4    &   4000             &   $0.52$       &    0.514               &      0.046               & 0.000  & 0.000  & 0.000  & 0.000  & 0.005  & 0.045        & 0.950    \\
 5    &   5000             &   $0.32$       &    0.783               &      0.101               & 0.000  & 0.000  & 0.000  & 0.000  & 0.005  & 0.045        & 0.950    \\
 6    &   6000             &   $0.12$       &    1.035               &      0.178               & 0.020  & 0.100  & 0.100  & 0.020  & 0.005  & 0.035        & 0.720    \\
 7    &   8000             &  $-0.18$       &    1.447               &      0.386               & 0.100  & 0.260  & 0.400  & 0.080  & 0.005  & 0.005        & 0.150    \\
 8    &  10000             &  $-0.58$       &    1.718               &      0.608               & 0.236  & 0.500  & 0.200  & 0.040  & 0.002  & 0.002        & 0.020    \\
 9    &  12000             &  $-0.98$       &    1.878               &      0.804               & 0.318  & 0.620  & 0.050  & 0.010  & 0.001  & 0.001        & 0.000    \\
10    &  14000             &  $-1.38$       &    1.967               &      0.955               & 0.393  & 0.600  & 0.005  & 0.001  & 0.000  & 0.001        & 0.000    \\
11    &  16000             &  $-1.78$       &    2.015               &      1.062               & 0.499  & 0.500  & 0.000  & 0.000  & 0.000  & 0.001        & 0.000    \\
12    &  18000             &  $-2.18$       &    2.039               &      1.132               & 0.499  & 0.500  & 0.000  & 0.000  & 0.000  & 0.001        & 0.000    \\
13    &  20000             &  $-2.58$       &    2.051               &      1.175               & 0.499  & 0.500  & 0.000  & 0.000  & 0.000  & 0.001        & 0.000    \\
14    &  24000             &  $-3.18$       &    2.059               &      1.214               & 0.499  & 0.500  & 0.000  & 0.000  & 0.000  & 0.001        & 0.000    \\
\hline
\end{tabular}
\\[1.5ex]
\flushleft
$^a$~Outer boundary of the shell (\kms).\quad
$^b$~$\rho$ in g\,cm$^{-3}$.\quad
$^c$~Cumulative mass in \Msun.\quad
$^d$~Cumulative kinetic energy in $10^{51}$ erg.
\end{center}
\end{footnotesize}
\end{table*}

\subsection{Putting the pieces of the puzzle together}
\label{Putting the pieces of the puzzle together}

In the previous section we have seen that the 09dc-exp model of \citet{hachinger2012a} 
significantly overestimates the radioactive tail with respect to the observations 
of SN~2009dc. Though this is just a single model, the consequences of this finding 
are far-reaching. This is because 09dc-exp is quite generic for models that try to 
explain the peak luminosity of superluminous SNe~Ia by radioactivity (and hence 
need $\gtrsim 1.5$ \Msun\ of \Nifs), and at the same time keep the kinetic energy 
per mass low [which requires a significant amount of unburned material and a total 
ejecta mass of $>2.5$ \Msun; see also \citet{taubenberger2011a}]. In these models 
there is no obvious way to avoid the strong $\gamma$-ray trapping seen in 09dc-exp. 
The consequence is that probably all models with excessively large \Nifs\ masses 
will fail. 

If this is true and the luminosity of superluminous SNe~Ia around peak does not 
solely come from radioactive decay, other processes have to contribute. Especially 
ejecta--CSM interaction has to be considered as a possibility, despite the known 
caveats that the CSM would have to be H- and He-free in order to avoid narrow lines 
of those elements in the spectra of superluminous SNe~Ia, and that even with a C/O 
CSM it is not evident that telltale spectroscopic signatures can be avoided. The 
09dc-int model of \citet{hachinger2012a}, as a first attempt in this direction, has 
too little flux during the tail phase when the interaction should have come to an end. 

The 09dc-tail model has been constructed so as to improve on this aspect and reproduce 
the luminosity of SN~2009dc during the tail phase between 60 and 180\,d after the 
explosion. It has been built with a physical picture in mind that is quite similar to 
that of 09dc-int: the kinetic energy of $1.2\times10^{51}$ erg is typical of a normal 
SN~Ia from a Chandrasekhar-mass WD progenitor. Also the \Nifs\ mass of 1.0 \Msun\ is 
within the range normally observed in SNe~Ia \citep{stritzinger2006b}, though on the 
upper end. While the 09dc-int model only represents the ejecta of a \MCh\ WD, the 
total ejecta mass of 09dc-tail is significantly larger. The additional mass can be 
interpreted as swept-up circumstellar material, leading to reduced ejecta velocities 
and higher densities compared to a freely expanding SN~Ia. At the same time, a small 
fraction of the kinetic energy would be transformed into light in the ejecta--CSM 
interaction, potentially compensating the flux deficit of the 09dc-tail model at 
early phases. Whether or not this may work and whether the resulting early-time 
spectra would resemble those of superluminous SNe~Ia can only be decided in detailed 
radiation-hydrodynamics simulations \citep[e.g.][]{fryer2010a,blinnikov2010a,noebauer2012a}, 
which go beyond the scope of this work. 

Nevertheless, a few estimates on the required CSM properties and a possible progenitor 
system can be made. The CSM would have to be H- and probably He-free, since these 
elements are not detected in the spectra of superluminous SNe~Ia at any phase. Most 
likely it would instead consist of a mix of C and O, which would be naturally 
explained if it had been produced in a merger of two C/O WDs, with the secondary 
being disrupted and its material being accreted slowly on to the primary. 
The masses of these two WDs would have to sum up to $\sim$\,2 \Msun. 
At the time the accreting primary approaches \MCh\ and explodes, it would be surrounded 
by 0.6--0.7 \Msun\ of CSM, distributed in an extended spherical envelope supported by 
thermal pressure \citep{shen2012a,schwab2012a}.\footnote{We emphasise that our 09dc-tail 
model has been constructed in a self-consistent way, in the sense that the kinetic 
energy of the ejecta agrees with the energy release by nuclear burning (see 
Table~\ref{09dc-blablub} for the composition) minus the binding energy of a \MCh\ WD.} 
This scenario is very similar to that 
studied by \citet{fryer2010a}, who found that in a SN~Ia enshrouded by a C/O-rich CSM 
originating from a merger event the light curve and spectra are altered substantially 
by shock heating. The light-curve peak tends to become much broader and a strong UV 
flux is generated, a feature that is also observed in early spectra of superluminous 
SNe~Ia \citep[see e.g.][]{scalzo2010a,silverman2011a,taubenberger2011a}.

To reconcile the synthetic bolometric light curve of the 09dc-tail model with 
observations, additional luminosity from interaction is required during the first 
$\sim$\,60\,d after the explosion (Fig.~\ref{fig:models}, bottom panel). This 
coincides with the duration of the peak phase, which may suggest that there is not 
ongoing CSM interaction over this full period of time, but that the photons are 
produced during an early and comparatively short interaction episode, trapped in 
the optically thick ejecta and released on photon-diffusion time scales 
\citep[see also][]{scalzo2012a}.

Having swept up 0.6--0.7 \Msun\ of material composed of up to 50 per cent of carbon, 
the ejecta of such an explosion would contain at least an order of magnitude more carbon 
than ordinary delayed detonations of a naked \MCh\ WD \citep{seitenzahl2013a}.
The strong and persistent \CII\ features in the early spectra of superluminous SNe~Ia 
would in this scenario not come as a surprise. Moreover, the same carbon might form 
dust at later phases, as already discussed in Section~\ref{Late-time dust formation?}, 
and thus explain the luminosity drop in at least some superluminous SNe~Ia during the 
nebular phase.

Of course, the proposed scenario is not free of problems. For example, it is not clear 
how much fine-tuning is necessary to produce a CSM with the right properties in a WD 
merger. As mentioned by \citet{fryer2010a}, besides the total mass also the density 
profile of the circumstellar material has an enormous impact on the resulting light 
curves and spectra. To explain observed superluminous SNe~Ia, one would need a 
configuration that boosts the light-curve peak for $\sim$\,60\,d, slows down the 
light-curve evolution, but modifies the spectra only moderately, preserving 
most characteristic SN~Ia features. Another critical point is that slow mergers 
of such massive WDs might not exist. The closer the mass ratio of the primary to the 
secondary WD is to unity, and the more massive the two WDs are, the more likely the 
merger will proceed violently \citep{pakmor2012a}, not leading to an extended C/O 
envelope. Binary population synthesis calculations do not predict primary WD masses 
above $1.3$ \Msun\ in C/O-WD mergers \citep{ruiter2013a}. To arrive at a system mass 
of $\sim$\,2 \Msun, both WDs hence need to be relatively massive. Finally, if 
non-violent mergers of C/O WDs do exist, the primary may collapse to a neutron star 
owing to electron captures rather than explode as a SN~Ia \citep{saio1985a,shen2012a}. 
Nonetheless, the proposed scenario of a slow merger of two massive C/O WDs that leads 
to a \MCh\ SN~Ia explosion enshrouded by a C/O-rich CSM remains a promising explanation 
for superluminous SNe~Ia. Its biggest strength is its ability to describe many thus 
far seemingly decoupled properties of superluminous SNe~Ia in a unified way as logical 
consequence of an enshrouded explosion, with no need to resort to unreasonable \Nifs\ 
and ejecta masses.

\section{Conclusions}
\label{Conclusions}

The late-phase photometry and spectra of a group of proposed super-Chandrasekhar 
SNe analysed in this work have highlighted interesting trends. The superluminous 
SNe~2006gz, 2007if and 2009dc are distinguished from normal and 91T-like SNe~Ia by 
nebular spectra with very weak [\FeIII] lines and a likely contribution of [\CaII] 
to the emission around 7300\,\AA. The ejecta of these objects are apparently not 
as highly ionised as in normal SNe~Ia, which could be a consequence of higher 
densities due to rather low expansion velocities. The proposed super-Chandrasekhar 
object SNF20080723-012 does not share these characteristics, and is instead very 
similar to SN~1991T in both its nebular spectra and its light-curve evolution. 
We are hence inclined to consider SNF20080723-012 as a classical 91T-like object 
rather than postulate fundamental diversity in the late-time behaviour of 
superluminous SNe~Ia. In fact, some diversity is present also in the late-time 
spectral energy distribution and light-curve decline of SNe~2006gz, 2007if and 
2009dc, with two of these objects showing a light-curve break after 
$\sim$\,150--200\,d leading to a more rapid fading thereafter. This, however, may 
be understood in terms of different time scales and intensities of dust formation.

Accepting this explanation, studying the light-curve tail before the possible 
break (i.e., at phases when the light curves are presumably powered by radioactive 
decay of \Cofs\ and opacities to optical photons are low) may be a promising way to 
constrain \Nifs\ and total ejecta masses of superluminous SNe~Ia, and thus get a 
handle on the nature of these objects. To this end we have compared the observed 
$U\!BV\!RIJHK$-bolometric light curve of SN~2009dc with synthetic bolometric light 
curves of models proposed in the literature. We find that models that have enough 
\Nifs\ to explain the light-curve peak by radioactive decay and at the same time 
enough mass to avoid too high ejecta velocities, are almost inevitably too luminous 
on the radioactive tail because of too strong $\gamma$-ray trapping. This 
dilemma may be overcome with models that assume additional luminosity from 
ejecta--CSM interaction during the peak phase. Improving on earlier work, we have 
presented one such toy model with $\sim$\,1 \Msun\ of \Nifs\ and $\sim$\,2 \Msun\ 
of ejecta that provides a convincing match to the observed light-curve tail of 
SN~2009dc. This model may be interpreted as a `normal' SN~Ia explosion of a \MCh\ 
WD enshrouded by 0.6--0.7 \Msun\ of C/O-rich material that is swept up as the 
ejecta expand. Such a configuration could be the outcome of a non-violent merger 
of two massive C/O WDs.

\section*{Acknowledgments}

We thank the anonymous referee for their comments that helped to improve 
the paper. We are also grateful to K. Kawabata and K. Maeda for providing us their 
nebular spectrum of SN~2006gz.
This work has been supported by the Transregional Collaborative Research Center 
TRR 33 `The Dark Universe' of the Deutsche Forschungs\-gemeinschaft, the Excellence 
Cluster EXC153 `Origin and Structure of the Universe', the PRIN-INAF 2009 with the 
project `Supernovae Variety and Nucleosynthesis Yields', the programme ASI-INAF 
I/009/10/0 and the ARCHES prize of the German Ministry of Education and Research 
(BMBF). It has benefited from data taken by the European Supernova Collaboration 
led by SB.
Observations were collected at the ESO 8.2\,m Very Large Telescope UT1 and UT2 
(Cerro Paranal, Chile, programmes 281.D-5043, 083.D-0728 and 085.D-0701) and the 
2.2\,m Telescope of the Centro Astron\'omico Hispano Alem\'an (Calar Alto, Spain). 
We thank the astronomers at both observatories for their support.
This research made use of the Weizmann Interactive Supernova data REPository 
(WISeREP; \citealt{yaron2012a}).

% \bibliography{astrofritz} \bibliographystyle{mn2e}

\addcontentsline{toc}{chapter}{Bibliography}
\markboth{Bibliography}{Bibliography}
\bibliographystyle{mn2e}

\appendix

\section{Tables}

\begin{table}
\caption{Magnitudes of the local sequence stars in the field of
SNF20080723-012 (Fig.~\ref{fig:chart}, top).} 
\label{std}
\begin{footnotesize}
\begin{center}
\begin{tabular}{@{}ccccc@{}}
\hline
ID & $B$ & $V$ & $R$ & $I$\\
\hline
 1  &  $20.97\pm0.03$  &  $20.49\pm0.02$  &  $20.16\pm0.02$  &  $19.84\pm0.02$\\
 2  &  $20.64\pm0.03$  &  $19.15\pm0.01$  &                  &                \\
 3  &  $21.53\pm0.02$  &  $21.01\pm0.03$  &  $20.68\pm0.03$  &  $20.34\pm0.02$\\
 4  &  $22.05\pm0.02$  &  $21.55\pm0.02$  &  $21.24\pm0.02$  &  $20.86\pm0.03$\\
 5  &  $21.55\pm0.03$  &  $20.54\pm0.02$  &  $19.87\pm0.03$  &  $19.36\pm0.05$\\
 6  &  $19.26\pm0.04$  &  $18.81\pm0.04$  &                  &                \\
 7  &  $19.52\pm0.02$  &  $18.92\pm0.02$  &                  &                \\
 8  &  $23.06\pm0.03$  &  $21.56\pm0.03$  &  $20.40\pm0.02$  &  $19.03\pm0.06$\\
 9  &  $22.66\pm0.03$  &  $21.72\pm0.02$  &  $21.10\pm0.04$  &  $20.55\pm0.03$\\
10  &  $22.59\pm0.04$  &  $22.09\pm0.02$  &  $21.80\pm0.05$  &  $21.42\pm0.06$\\
11  &  $22.00\pm0.04$  &  $20.56\pm0.02$  &  $19.57\pm0.02$  &                \\
12  &  $20.32\pm0.04$  &  $19.28\pm0.01$  &                  &                \\
\hline\\[-0.7ex]
\end{tabular}
\end{center}
\end{footnotesize}
\end{table}

\begin{table*}
\caption{$S$- and $K$-corrections applied to the photometry of SN~2007if and SNF20080723-012.}
\label{SKcorr} 
\begin{footnotesize}
\begin{center}
\begin{tabular}{@{}lcrrrrrrrrrl@{}}
\hline
MJD       & SN    & Epoch$^a$ & $S_B\ $ & $S_V\ $ & $S_R\ $ & $S_I\ $ & $K_B\ $ & $K_V\ $ & $K_R\ $ & $K_I\ $ & Telescope \\
\hline
54\,733.2 & SN~2007if & 358.2 &         &$-0.038$ &         &         &         &$ 0.322$ &         &         & VLT \\
54\,739.8 & SNF       &  54.8 &$ 0.042$ &$-0.051$ &$-0.004$ &         &$-0.261$ &$-0.161$ &$ 0.102$ &         & CAHA \\
54\,882.2 & SNF       & 187.3 &$ 0.051$ &$ 0.016$ &$ 0.009$ &         &$-0.481$ &$ 0.273$ &$ 0.223$ &         & CAHA \\
54\,947.4 & SNF       & 248.0 &$-0.087$ &$-0.086$ &$-0.198$ &$-0.004$ &$-0.592$ &$ 0.490$ &$ 0.283$ &$ 0.058$ & VLT \\
54\,973.2 & SNF       & 272.0 &$-0.087$ &$-0.086$ &$-0.198$ &$-0.004$ &$-0.592$ &$ 0.490$ &$ 0.283$ &$ 0.058$ & VLT \\
\hline
\end{tabular}
\\[1.5ex]
\flushleft
$^a$~Phase in rest-frame days with respect to $B$-band maximum [MJD $= 54\,348.4$ 
     for SN~2007if \citep{scalzo2010a} and MJD $= 54\,680.9$ for SNF20080723-012 
     \citep{scalzo2012a}].
\end{center}
\end{footnotesize}
\end{table*}

\label{lastpage}

\end{document}